\documentclass[fleqn,usenatbib]{mnras}
\usepackage{dcolumn}
\usepackage{bm}
\usepackage{xcolor}
\usepackage{enumitem}[labelwidth=*]
\usepackage{hyperref}
\usepackage{verbatim}
\usepackage{amsmath,amstext,mathrsfs}
\usepackage[all]{hypcap} 
\usepackage{afterpage}    
\usepackage{marginnote}
\usepackage{makecell}
\usepackage{subfigure}
\usepackage{xfrac}
\usepackage{soul}
\usepackage{float}

\usepackage{ifxetex,ifluatex}
\usepackage{multirow}
\usepackage{amsfonts}
\usepackage{amssymb}

\DeclareFontFamily{OMS}{oasy}{\skewchar\font48 }
\DeclareFontShape{OMS}{oasy}{m}{n}{%
         <-5.5> oasy5     <5.5-6.5> oasy6
      <6.5-7.5> oasy7     <7.5-8.5> oasy8
      <8.5-9.5> oasy9     <9.5->  oasy10
      }{}
\DeclareFontShape{OMS}{oasy}{b}{n}{%
       <-6> oabsy5
      <6-8> oabsy7
      <8->  oabsy10
      }{}
\DeclareSymbolFont{oasy}{OMS}{oasy}{m}{n}
\SetSymbolFont{oasy}{bold}{OMS}{oasy}{b}{n}

\DeclareMathSymbol{\smallleftarrow}     {\mathrel}{oasy}{"20}
\DeclareMathSymbol{\smallrightarrow}    {\mathrel}{oasy}{"21}
\DeclareMathSymbol{\smallleftrightarrow}{\mathrel}{oasy}{"24}

\newcommand*{\kh}{} 
\newcommand*{\qq}{\color{purple}\bf} 
\newcommand*{\hy}{}
\graphicspath{{./}{Fig/}}

\newcommand*{\torefereetwo}{}

\title[Alfven leakage]{Anomalous compressible mode generation by global frame projections of pure Alfven mode}
\author[Yuen et.al]{
Ka Ho Yuen,$^{1,2}$\thanks{kyuen@lanl.gov (Oppenheimer Fellow), ORCID: 0000-0003-1683-9153}
Huirong Yan,$^{3,4}$\thanks{huirong.yan@desy.de, ORCID: 0000-0003-2560-8066}
Alex Lazarian$^{1}$\thanks{alazarian@facstaff.astro.wisc.edu}
\\
$^{1}$Department of Astronomy, University of Wisconsin-Madison, USA, 53715\\
$^{2}$Theoretical Division, Los Alamos National Laboratory, Los Alamos, NM 87545, USA\\
$^{3}$Deutsches Elektronen-Synchrotron DESY \\
$^{4}$Institut für Physik \& Astronomie, Universität Potsdam, Germany
}

\date{Accepted XXX. Received YYY; in original form ZZZ}
\begin{document}

\label{firstpage}
\pagerange{\pageref{firstpage}--\pageref{lastpage}}
\maketitle

\pubyear{2022}
\begin{abstract}
{
Alfven wave is the single most important physical phenomenon of magneto-hydrodynamic turbulence and has far-reaching impact to almost all studies related to astrophysical magnetic field.
Yet the restoration of the Alfven wave fluctuations from a given magnetic field, aka the local Alfven wave problem, is never properly addressed in literature albeit its importance. 
Previous works model the Alfven wave fluctuation as the perturbation along a straight-line, constant magnetic field. However, {\torefereetwo Lazarian \& Pogosyan (2012)} suggested that the decomposition of Alfven wave along a straight line, aka. the global frame decomposition, has a factor of discrepancy to the true local Alfven wave fluctuation. 
Here we provide a geometric interpretation on how the local Alfven wave is related to the global frame through the use of vector frame formulation. We prove both analytically and numerically that the local frame Alfven wave is an orthogonal transformation of that of the global frame and related by the local Alfvenic Mach number. In other words, when we observe Alfven wave in the global frame of reference, some of the Alfven wave will be mistaken as compressible waves.
The importance of frame choices have a far-reaching impact to the analytical studies of MHD turbulence. Combining the frame formalism and the new techniques we can have accurate measurement to some of the fundamental turbulence properties like the inclination angle of mean magnetic field relative to the line of sight.
}


\end{abstract}


\section{Introduction} \label{sec:intro}
Turbulence is ubiquitous in astrophysical environment and the interstellar gases are permeated by turbulent magnetic fields. Magneto-hydrodynamic (MHD) turbulence plays a very important role in various astrophysical phenomena (see \cite{1995ApJ...443..209A, 2010ApJ...714.1398C,2003matu.book.....B,2004ARA&A..42..211E,2007ARA&A..45..565M}), including star formation (see \cite{2007ARA&A..45..565M,2016ApJ...824..134F}), propagation and acceleration of cosmic rays (see \cite{2000PhRvL..85.4656C,2002PhRvL..89B1102Y,2004ApJ...604..671F,2016ApJ...833..131L}), as well as regulating heat and mass transport between different ISM phases (\cite{1993MNRAS.262..327G,  2000ApJ...543..227D,2001ApJ...561..264D,2004ApJ...616..943L,2006ApJ...652.1348L,2006ApJS..165..512K,2006MNRAS.372L..33B,2006ApJ...653L.125P} see \cite{2009ASPC..414..453D,2011piim.book.....D} for the list of the phases).


MHD turbulence is usually highly compressible, and has been thoughtfully studied by a number of authors in the community (e.g. \cite{2007ApJ...658..423K}). However, the compressibility of the turbulence adds additional difficulty in the understanding of how the three fundamental MHD modes (namely Alfven, slow and fast modes) would behave in various astrophysical phenomena, each carrying different spectra and anisotropies. For instance, it is believed that the Alfven mode plays a central role in making the cold neutral media aligned with the magnetic field \citep{2018ApJ...865...46L} {\hy and controls the transport of heat and particles across magnetic fields \citep{2001ApJ...562L.129N,2006ApJ...645L..25L,2008ApJ...673..942Y,2021arXiv210801936M}}. In comparison, fast modes play an important role in the scattering and acceleration of cosmic rays (\citealt{2002PhRvL..89B1102Y,2004ApJ...614..757Y,2005ThCFD..19..127C,2008ApJ...686..350L,2007MNRAS.378..245B}). The modes composition strongly depends on the way of driving \cite{2020PhRvX..10c1021M}. It is therefore essential to have a handy way in decomposing the three fundamental MHD modes in numerical analysis.

A notable development is the statistical mode decomposition developed by \citeauthor{2002PhRvL..88x5001C} (\citeyear{2002PhRvL..88x5001C,2003MNRAS.345..325C}, latter hereafter CL03), which allows one to obtain the realization of the three fundamental MHD modes in numerical simulations by considering a perturbation along a locally strong magnetic field direction. The realization of MHD modes allowed the community to validate the theory of MHD turbulence (\cite{1995ApJ...438..763G} hereafter GS95, see also \cite{1999ApJ...517..700L,2000ApJ...539..273C,2001ApJ...554.1175M,2001AAS...198.9003L,2002PhRvL..88x5001C,2003MNRAS.345..325C}) through numerical simulations. In particular, the scaling relation of compressible modes were first verified through the realization of MHD modes using the mode decomposition algorithm developed by CL03. The realization of MHD modes also excites the development of different techniques in studying MHD turbulence in observations, including the Velocity Gradient Technique (VGT, \cite{2017ApJ...837L..24Y,2017arXiv170303026Y}) which uses the anisotropy of different modes in retrieving the magnetic field directions in spectroscopic data, and also the Synchrotron Polarization Analysis (SPA, \cite{2020NatAs...4.1001Z}) which utilizes the properties of the projected statistics in predicting the dominance of Alfven or compressible modes in observational synchrotron data, as well as detailed analysis of solar wind turbulence \cite[e.g.][]{2021ApJ...923..253Z,Zhao22}.

However, \cite{1995ApJ...438..763G} model of MHD turbulence is of centre importance in the modern theory of MHD turbulence. The latter is employs the concept of "local frame of reference" that was added to the theory later \citep{1999ApJ...517..700L,2000ApJ...539..273C}.  This means that the eddies, which are usually elliptical in shape, are aligned to the local magnetic field rather than the mean magnetic field. As pointed out by \cite{2010ApJ...720..742K}, the decomposition of CL03 is a global frame decomposition, as opposed to the local frame MHD theory stressed in the works that followed the original GS95 study \citep{1999ApJ...517..700L}. As described in Fig.\ref{fig:mode_decomp}, when one considers a different volume, the realization of the three fundamental modes will be different due to the change of the mean magnetic field directions under the CL03 decomposition algorithm. The difficulty of obtaining the statistics of three modes in a localized manner has been attempted, including abandoning the realization of modes but focusing on the structure functions \cite{2005ApJ...624L..93B}, decomposing the MHD quantities into linear combination of fundamental localized patches before performing the CL03 decomposition \citep{2010ApJ...720..742K}, or the introduction of the frame changing parameters in the framework of correlation functions \citep{LP12}.  Yet, how the three fundamental modes are realized in the local system of reference is still an unsolved problem for numerical community.


In this paper, we explore how the Alfven and compressible modes in the local system of reference are expressed globally. In \S \ref{sec:mode} we review the CL03 method and its possible improvements. In particular, in \S \ref{subsec:leakage} we discuss about the generation of "compressible waves signature" due to the wrong choice of local frame of reference. From \S \ref{sec:SPA} to \S \ref{sec:gamma}, we describe a few applications that utilize the concept of Alfven leakage, namely the applications of the Synchrotron Polarization Analysis Technique to regimes with strong Faraday rotation (\S \ref{sec:SPA}) and the determination of the line of sight angle $\gamma$ (\S \ref{sec:gamma}).  In \S \ref{sec:discussion} we discuss about the possible impacts of our method and the caveats of our work. In \S \ref{sec:conclusion} we conclude our paper.

\section{Mode decomposition}
\label{sec:mode}
\subsection{Review of the MHD mode decomposition methods}
\begin{figure*}
\includegraphics[width=0.98\textwidth]{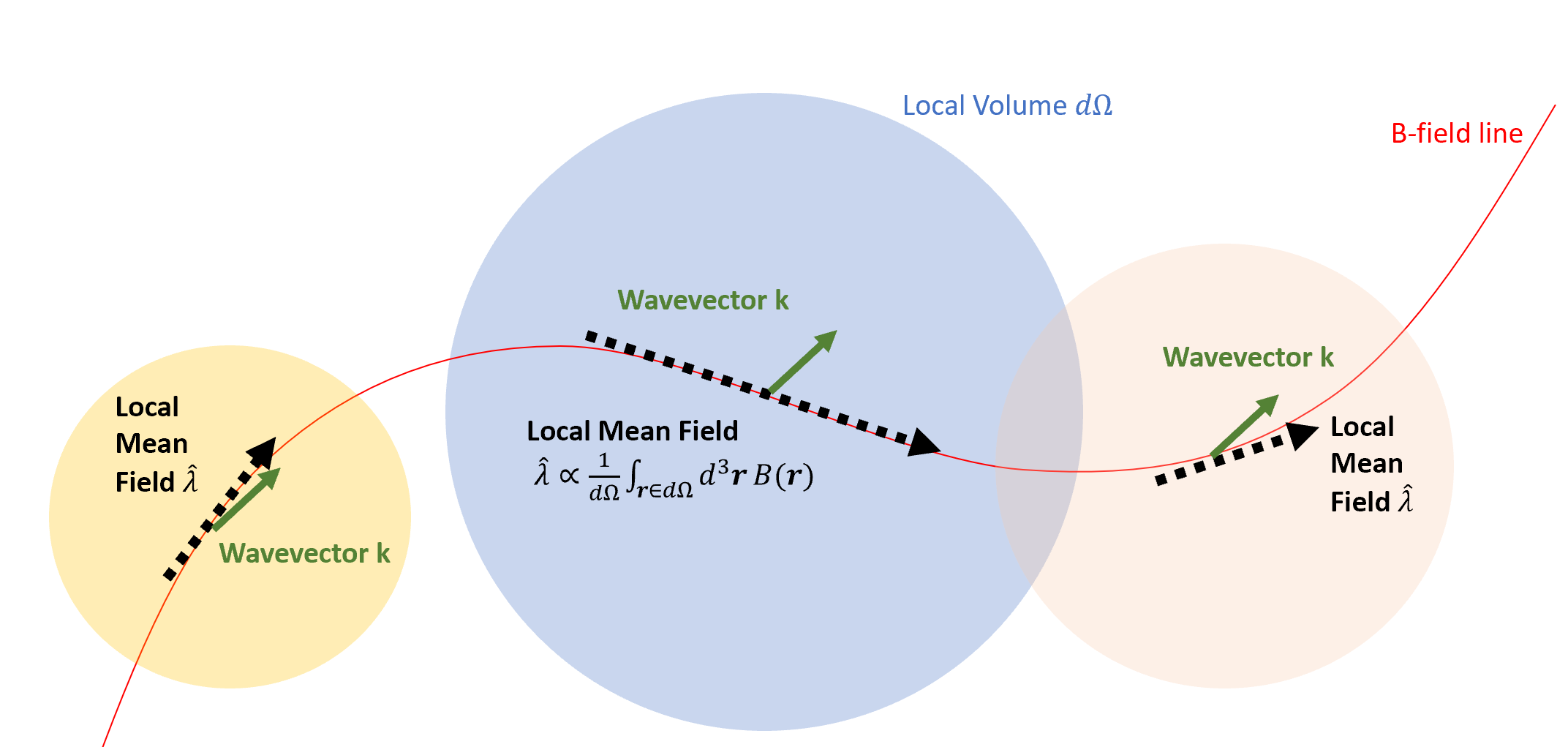}
\caption{The concept of mode decomposition in CL03: (1) By selecting a volume $d\Omega$, a local mean magnetic field direction $\hat{\lambda}$ would then be defined for later decomposition. (2) All the wavevector ${\bf k}$ that is contained in this volume $d\Omega$ are used for the decomposition. (3) For each ${\bf k}$ there is a local reference frame (See Fig.\ref{fig:illus}) that decomposes the magnetic field into the three eigenmodes. The change of the selected volume $d\Omega$ will lead to different mean field vector $\hat{\lambda}$. As a result, the local decomposition result would be functions of both $\lambda$ and the wavevector ${\bf k}$ .}
\label{fig:mode_decomp}
\end{figure*}

In this section we review the underlying assumptions of the mode decomposition method as introduced by CL03 and the development since then. In CL03 they consider a volume $d\Omega$ in which the perturbation of magnetic field is small compared to the mean field $\delta B(d\Omega)< \langle B\rangle$, so does the density fluctuations $\delta \rho/\langle \rho\rangle < 1$. Fig.\ref{fig:mode_decomp} shows how the volume $d\Omega$ is selected. Readers should be careful that once the volume is selected the  mean magnetic field direction $\hat{\lambda}$ is also defined respectively. In this scenario,  the small perturbation in the presence of a strong mean magnetic field will provide a linearized set of MHD equations in which three non-trivial eigenvectors would be found. In this localized box, the Alfven, slow and fast mode eigenvectors are\footnote{However, recent publication from \cite{4DFFT} suggests that a significant portion of the projected spectral powers are not in the form of propagating waves, but fluctuations with miniature frequencies. The nature of the {\torefereetwo non-wave fluctuations} as dubbed in \cite{4DFFT} requires further clarifications from the theory of MHD turbulence. See \cite{2019tuma.book.....B,2022ApJ...936..127F,2022JPlPh..88e1501S}.}:
\begin{equation}
    \begin{aligned}
    \zeta_A(\hat{\bf k},\hat{\lambda}) &\propto \hat{\bf k} \times \hat{\bf  \lambda}\\
    \zeta_S(\hat{\bf k},\hat{\lambda}) &\propto (-1 +\alpha-\sqrt{D}) ({\bf k}\cdot \hat{\bf \lambda}) \hat{\bf \lambda}  \\&+ (1+\alpha - \sqrt{D}) ( \hat{\bf \lambda} \times ({\bf k}\times \hat{\bf \lambda})) \\
    \zeta_F(\hat{\bf k},\hat{\lambda}) &\propto (-1 +\alpha+\sqrt{D}) ({\bf k}\cdot \hat{\bf \lambda}) \hat{\bf \lambda}  \\&+ (1+\alpha + \sqrt{D}) ( \hat{\bf \lambda} \times ({\bf k}\times \hat{\bf \lambda})) \\
    \end{aligned}
    \label{eq:cho}
\end{equation}
where $\alpha = \beta\gamma/2$, $D=(1+\alpha)^2- 4\alpha\cos ^2\theta_\lambda$, $\cos\theta_\lambda = \hat{\bf k}\cdot \hat{\bf \lambda}$, plasma $\beta\equiv P_{gas}/P_{mag}$ measures the compressibility and $\gamma =\partial P/\partial \rho$ is the polytropic index of the adiabatic equation of state. The presence of $\hat{\bf k}$ suggests that the direction of the three mode vectors are changing as ${\bf k}$ changes. In this scenario, the perturbed quantities, say for the velocity fluctuations ${\bf v}_1 = {\bf v}-\langle {\bf v}\rangle$ can be written as:
\begin{equation}
    {\bf v}_1({\bf r}) = \int d^3 {\bf k} e^{i{\bf k}\cdot {\bf r}} \sum_{X\in A,S,F} F_{0,X}({\bf k})F_{1,X}({\bf k},\hat{\lambda})  C_X \zeta_X(\hat{\bf k},\hat{\lambda})
    \label{eq:2}
\end{equation}
for some power spectrum $E_v(k) = F^2_0 = k^{-n/2}$ \citep{spectrum}, some anisotropy weighting function $F_1$ and the mode constants $C_X$ denoting the relative weight of the three modes. Notice that the decomposition (Eq.\eqref{eq:cho}) is a {\bf global} decomposition method since the magnetic field fluctuations {\it within the volume $d\Omega$ is not considered} when computing the eigenvectors of the three modes. One of the most important consequences of performing global decomposition is the loss of the GS95 scaling for small $k$. In fact, \cite{2002PhRvL..88x5001C} (see also \cite{LP12}) pointed out that in the global system of reference the anisotropic scaling is scale independent, meaning that the elongation of turbulence eddies is fixed and does not change as the eddies cascade.  Another important consequence of the concept of $d\Omega$ is that, when one changes the sampling volume, e.g. from the volume in blue to that of yellow or orange in Fig.\ref{fig:mode_decomp}, the weighting of the three modes will also change due to the change of the mean direction of magnetic field from volume to volume. 

Notice that the selection of the volume $d\Omega$ has to fulfill the conditions as assumed in CL03: Both the sonic and Alfven Mach number within the volume must be smaller than unity. Notice that if the volume is smaller than the volume defined by the correlation length of the turbulence, the dispersion of the turbulence observables will be scaled as a function of distance according to their respective turbulence statistics, meaning that if $\rho,v$ follows GS95, 
\begin{equation}
    \begin{aligned}
    \delta \rho^2({\bf r}) &\propto r^{2/3}\\
    \delta v^2({\bf r}) &\propto r^{2/3}
    \end{aligned}
\end{equation}
To address the issue of the locality, the community has explored a number of ways to include the local fluctuations of magnetic field during the calculation of statistics of MHD modes. For instance, one of the most notable ways of obtaining the statistics of MHD modes is to compute the local structure functions \cite{2005ApJ...624L..93B}. The mathematical expression of the 3D structure function  of the turbulence variable $v$ in the local frame of reference is given by:
\begin{equation}
    SF\{{\bf v}\}({\bf r}) = \Big\langle \left(({\bf v}({\bf r'}+{\bf r})-{\bf v}({\bf r'}))\cdot \hat{\lambda}({\bf r},{\bf r'})\right)^2\Big\rangle_{\bf r'}
    \label{eq:localSF}
\end{equation}
where
\begin{equation}
    \hat{\lambda}({\bf r},{\bf r'})=\frac{{\bf B}({\bf r'}+{\bf r})+{\bf B}({\bf r'})}{|{\bf B}({\bf r'}+{\bf r})+{\bf B}({\bf r'})|}
\end{equation}
The anisotropy computed throughout this manner is scale dependent and exhibit the GS95 scaling $r_{\parallel} \propto r_{\perp}^{2/3}$. The use of the local structure function correctly recovers the GS95 statistics, yet it is not possible to obtain the realization of the modes in this manner, meaning that further study of the features of the modes, e.g. how does the mode look like when projected, are prohibited when using the structure functions. 

Another important way of improving the CL03 is to rewrite the turbulence variables into the linear sum of small localized patches through the wavelet transform \citep{2010ApJ...720..742K}. Physically, they are looking for specific functional forms obeying the orthogonality requirement and represent the volumes as shown in Fig.\ref{fig:mode_decomp}.By considering the set of orthogonal wavelets $\phi$, one can write the velocity field ${\bf v}({\bf r})$ as:
\begin{equation}
    \tilde{\bf v}({\bf w};a) = a^{-3/2} \int d^3 x \phi(\frac{{\bf r}-{\bf w}}{a}) {\bf v}({\bf r})
\end{equation}
where $\phi$ is the set of wavelet functions. For \citep{2010ApJ...720..742K} they select the 12-tap Daubechies wavelet. To perform mode decomposition, they first convert the velocity field into the linear combinations of wavelets, and then proceed with the procedures of CL03 for the wavelet transformed variable. The contributions for all wavelets for a specific mode are added before the inverse Fourier transform takes place.  Notice that the wavelet functions is simply a mathematical construction that may contain spatially dispatched regions (e.g. D4 and D12 of the Daubechies wavelet), for which taking the statistical calculations within the wavelet might not be physically justifiable. Nevertheless, the improved mode decomposition method proposed by \cite{2010ApJ...720..742K} still retrieve seemingly the correct GS95 statistics from the simulations.

\subsection{The locality of the mode decomposition problem due to magnetic field wandering effect}
\begin{figure*}
\includegraphics[width=0.98\textwidth]{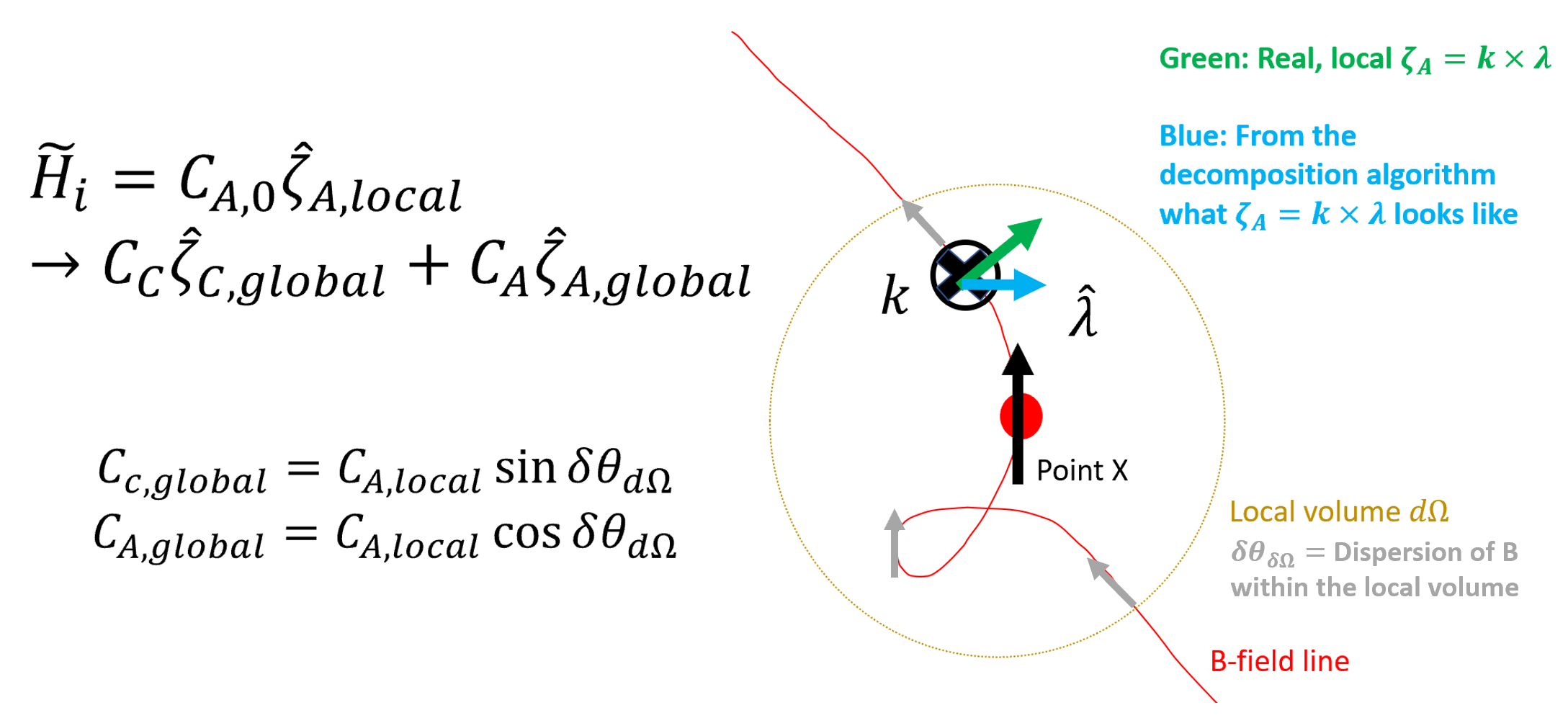}
\caption{An illustrative figure showing how the {\it Alfven Leakage} phenomenon happens during the mode decomposition process. Here the red line represents the B-field line, point X is the origin of the volume $d\Omega$, that is represented by the dash orange circle. The vector $\hat{\lambda}$ represents the mean field averaged over $d\Omega$. For a given point within $d\Omega$, assuming the k-vector points inside the plane, the local Alfven wave unit vector (green) makes an angle to that of the Alfven wave unit vector defined by $\hat{\lambda}$. This effect will be stronger if the magnetic field fluctuations within $d\Omega$ is larger, and vice versa.  }
\label{fig:alfleak}
\end{figure*}

In realistic MHD simulations the magnetic field lines are fluctuating within any selected volume, which is named wandering effect as in CL03. However the mode decomposition algorithm available in the community had not considered any of these wandering effect, which makes the estimation of modes be rather unrealistic for larger $M_A$. To model the additional effect when decomposing the modes in the global frame of reference, \citeauthor{LP12} (\citeyear{LP12}) model the {\it magnetic field} correlation function as the linear combination of the isotropic tensor $\hat{T}_E$ and the axis-symmetric tensor $\hat{T}_F$ (See \citealt{2004ApJ...614..757Y} also our Appendix for a detailed mathematical construction). In general the direct tensor of the magnetic field in the Fourier space at a given wavevector ${\bf k}$ in the local frame of reference can be written as:
\begin{equation}
    \tilde{H}_i\tilde{H}_j \propto E({\bf k}) \hat{T}_{E,ij} + F({\bf k}) \hat{T}_{F,ij}
\end{equation}
The transformation from the local to global frame as modelled by LP12 is:
\begin{equation}
\begin{aligned}
    \hat{T}_{E,local} &\rightarrow \hat{T}_{E,global}\\
    \hat{T}_{F,local} &\rightarrow W_I\hat{T}_{E,global} + W_L\hat{T}_{F,global}\\
\end{aligned}
\label{eq:LP12_WIWL}
\end{equation}
where $W_{I,L}$ are two modelling constants that are functions of $M_A$ and also ${\bf k}$. 

\section{Alfven Leakage}
\label{subsec:leakage}

\subsection{Description of the problem}

Since $\hat{T}_E = \hat{T}_C + \hat{T}_A$ and $\hat{T}_F=\hat{T}_C$ (See the Appendix \ref{subsec:T}) one can actually write the magnetic field in the Fourier space in the global frame of reference using vector notations due to the orthogonality of the base vectors:
\begin{equation}
    \tilde{H}({\bf k}) = C_C \zeta_C  + C_A\zeta_A
    \label{eq:vecnot}
\end{equation}
where 
\begin{equation}
\begin{aligned}
\zeta_C &= \frac{(\hat{\bf k}\times (\hat{\bf k} \times \hat{\bf  \lambda}))_i}{|\hat{\bf k} \times \hat{\bf  \lambda}|}\\
\zeta_A &= \frac{(\hat{\bf k} \times \hat{\bf  \lambda})_i}{|\hat{\bf k} \times \hat{\bf  \lambda}|}
\end{aligned}
\label{eq:frame_definition}
\end{equation}(See Appendix). The modelling of LP12 simply means that, assuming if in the local frame of reference the magnetic field only has the Alfven component $\tilde{H}({\bf k}) = \sqrt{E({\bf k})} C_0\zeta_A$, then the transformation from local to global reference frame simply means a vector projection:
\begin{equation}
    \tilde{H}({\bf k}) = \sqrt{E({\bf k})} C_0\zeta_A \xrightarrow{\text{global}} \sqrt{E({\bf k})}(C_C \zeta_C  + C_A\zeta_A)
\end{equation}
where $C_C = C_0 \sin(\theta_f)$, $C_A=C_0 \cos(\theta_f)$ for some $\theta_f$ that we will explore in the coming subsection. For Alfven waves, $E({\bf k)} = -F({\bf k})$, and the relation between $C_{C,A}$ and $W_{I,L}$ can be easily derived:
\begin{equation}
    \begin{aligned}
    C_A^2&=1-W_I\\
    C_C^2&=1-W_I-W_L
    \end{aligned}
\end{equation}
Notice that when $M_A \ll 1$, there is no magnetic field wandering effect. In this limit, $C_A(M_A\rightarrow 0)=1$, $C_C(M_A\rightarrow 0)=0$.

The vector notation (cf. Eq.\eqref{eq:vecnot}) allows us to think of this problem {\it geometrically} and relates to a very important physical effect that mentioned in both \cite{2018ApJ...865...54Y} and more recently \cite{2020ApJ...898...66Y}: The {\bf Alfven leakage} effect. Alfven leakage describes the effect that the locally Alfven component of the turbulent variables are projected as linear combinations of Alfven and non-Alfven components in the presence of a curved magnetic field { averaged over} selected volume ({\bf local mean magnetic field}). In \cite{2020ApJ...898...66Y} we consider the effect that the gravitational forces creates extra compression to the Alfven waves and thus some of the Alfven waves from the self-gravitating systems are transferred into compressible components. In fact, in the presence of non-trivial magnetic field structures, there exist non-zero \citep{2020ApJ...898...66Y} field lines in any volume due to magnetic field wandering. Similar effect has been considered in \cite{2001ApJ...554.1175M} but being corrected by \cite{2000ApJ...539..273C} as the "rotation effect" by recognizing that the global frame anisotropy axis ratio is a function of $M_A$. However the actual relation between the local and global frame of reference has yet to be explored.

Fig.\ref{fig:alfleak} gives a pictorial illustration on how the Alfven leakage happens during mode decomposition. For a given magnetic field $H_i({\bf r})$, its Fourier transform $\tilde{H}_i({\bf k})$ can be written as the linear sum of Alfven and compressible components. To simplify our argument, we consider a pure Alfven wave B-field in the local frame of reference, which $\tilde{H}_i$ is simply proportional to the local Alfven vector $\zeta_{A,local}$, which is represented by the green vector. However if the volume $d\Omega$ is selected (the orange dash circle), the mean field is defined in $d\Omega$. In this case, if we consider a pure Alfven mode magnetic field in the local frame of reference as we show in Fig.\ref{fig:alfleak}, the projection of the Alfven mode in the global frame of reference will be an linear combination of Alfven and compressible modes as written in Fig.\ref{fig:alfleak}. By selecting a k-vector that points inside the plane, the local Alfven wave unit vector (green) , which is defined by the cross product between the k-vector and the {\it local} magnetic field direction, makes an angle to that of the Alfven wave unit vector defined by $\hat{k}\times\hat{\lambda}$ which we will name that angle $\delta \theta_{d\Omega}$. The projection effect causes artificial compressible modes to be detected within $d\Omega$. It is very apparent that if the magnetic field line is aligned with the mean field vector $\hat{\lambda}$ in $d\Omega$, then there is no artificial compressible modes due to the projection effect. We call this effect {\it Alfven Leakage}. This effect is artificial and does not involve the change of the cascade laws or anisotropies.

\subsection{Modelling of the Alfven leakage problem}

We can further model the leakage phenomenon through the use of $\delta \theta_{d\Omega}$. From Fig.\ref{fig:alfleak} we see that $\delta \theta_{d\Omega}$ is a measure of the {\it average} {\torefereetwo magnetic field} angle dispersion in this selected volume. We can postulate that the dispersion of angles are related to the Alfven Mach number measured within $d\Omega$:
\begin{equation}
    \delta \theta_{d\Omega} \sim M_{A,d\Omega} 
    \label{eq:postulate}
\end{equation}
{\torefereetwo Notice that the dispersion of magnetic field angle can scale up to $M_A\approx 2$ as shown in the appendix of \cite{ch5}.} Notice that according to the definition of the Alfven Mach number and the self-similarity of turbulence cascade, any localized calculation of statistical observables in a turbulence system can be approximated by the scale relation through the observable's structure function. For the case of Alfven Mach number, the corresponding observable is $\delta B/B$, then 
\begin{equation}
\begin{aligned}
    M_{A,d\Omega} &\approx \left(\frac{1}{2B^2}\langle \delta B({\bf r}+{\bf r}')-\delta B({\bf r}')\rangle_{\bf r'}\right)^{1/2}\\
    &\sim M_{A,global} \left(\frac{r}{L_{inj}}\right)^{\kh 1/3}
\end{aligned}
\label{eq:scalingrel}
\end{equation}
where the last equality comes from the fact that the magnetic field fluctuation is small and Kolmogorov: $\langle \delta B^2\rangle \sim r^{\kh 2/3}$. We can then relate the $C_{C,A}$ with Eq.\eqref{eq:postulate} and \eqref{eq:scalingrel}.

\subsection{Numerical tests}

\begin{figure*}
\includegraphics[width=0.98\textwidth]{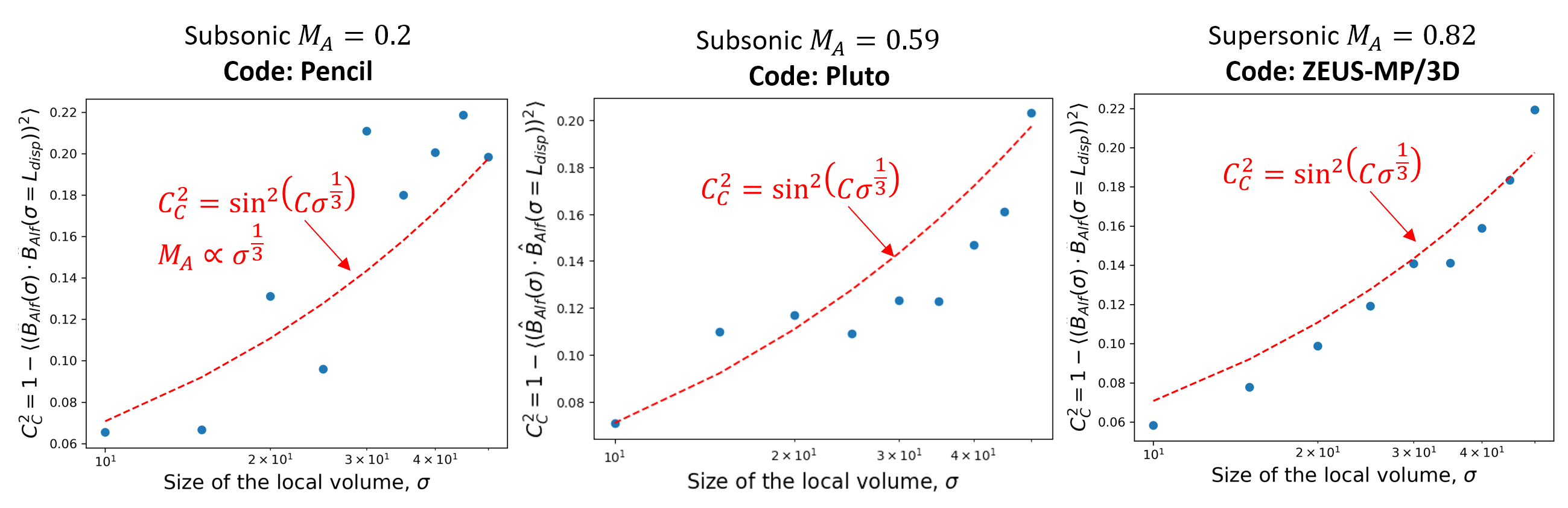}
\caption{Three figures showing the ratio of compressible modes $C_C$ as a function of the size of the volume selected $\sigma$ for three simulations. From the left: ma20, ma59, huge-4. (See Table \ref{tab:sim}).}
\label{fig:alfleak_incomp}
\end{figure*}

{\kh We can perform a very simple numerical test to illustrate the behavior of the Alfven leakage, which can be done by the following steps:
\begin{enumerate}

    \item We select the local frame vectors $\zeta_{A,C}$, where they can be {\it approximated} by selecting a very small volume in the numerical simulations and use Eq.\ref{eq:vecnot}. We shall call this volume 1.
    \item We then perform Alfven wave decomposition using $\zeta_A$ from simulations so that we can put $C_{A,local}=1$ in all our case (See Fig.\ref{fig:mode_decomp}).  
    \item We then select a larger volume with size $\sigma$. According to \S \ref{sec:mode}, if there is indeed the effect of Alfven leakage happening, then $C_C = \sin M_A \sim \sin (C \sigma^{1/3})$, where $C = M_{A,global}L^{-1/3}$.  In this new volume, there will be another pair of $\zeta_{A,C}$ being defined. We shall call this volume 2.
    \item We plot the quantity $C_C^2 = 1- (\langle \hat{B}_{Alf,1}\cdot \hat{B}_{Alf,2})^2\rangle$. Notice we only compare the directions of $B$ within volume 1, since this calculation only makes sense there.
    \item There could be three outcome from this test
    \begin{enumerate}
        \item If there is no such leakage effect, $C_C$ should be a constant zero as we already removed all non-compressible components in the previous step.
        \item If there is indeed $C_C$ but our model in the previous section is incorrect, then there should not be a dependence of $C_C\sim \sin (C\sigma^{1/3})$.
        \item If the Alfven leakage effect indeed exists, we expect $C_C\sim \sin (C\sigma^{1/3})$.
    \end{enumerate}
\end{enumerate}
Fig.\ref{fig:alfleak_incomp} shows how the $C_{C}$ behave as a function of the size of the volume $\sigma$, where we plot the regime when $\sigma$ is not in the dissipation range. Notice that here we intentionally pick simulation cubes from various numerical codes and with different conditions to show that the leakage effect is universal and rather independent to what choices of MHD code one works with. For each figure, we draw the predicted proportionality $C_C \sim \sin^2(C\sigma^{1/3})$ as the red dash curve. As we can see from these three subplots, our prediction $C_C \sim \sin^2(C\sigma^{1/3})$, which came analytically from previous sections and not from a fitting algorithm,  follows the trend reasonably well. 
}

This indicates that (1) the Alfven leakage effect actually exists even in incompressible mode turbulence (2) the leakage is smaller when one goes to smaller scales (3) our postulate that $\delta \theta_{d\Omega}\sim M_{A,local}$ is a good approximation. These three consequences indicate that the mode decomposition as proposed by CL03 requires an additional update in addressing the contributions of large scale magnetic field wandering to the relative composition of modes within the volume.

As a direct consequence of this section, the $W_{I,L}$ constants originated from \cite{LP12} as a function of $M_{A,local}$ ($\ll 1$) are given by Eq.\eqref{eq:LP12_WIWL} are :
\begin{equation}
    \begin{aligned}
    W_I &= {1-C_A^2 \propto M_A^2} \\
    W_L &= {C_A^2-C_C^2 \propto 1-2M_A^2}
    \label{eq:our_WIWL}
    \end{aligned}
\end{equation}

\begin{table}
\centering
\begin{tabular}{c c c c c}
Model & $M_S$ & $M_A$ & $\beta=2M_A^2/M_S^2$ & Resolution \\ \hline \hline
\multicolumn{5}{l}{\bf ZEUS-MP Simulations}\\
Ms0.92Ma0.09             & 0.92  & 0.09 & 0.02 & $480^3$ \\
Ms0.98Ma0.32            & 0.98  & 0.32 & 0.22 & $480^3$ \\
Ms0.93Ma0.94              & 0.93  & 0.94 & 2.0 & $480^3$ \\ 
huge-0                  & 6.17  & 0.22 & 0.0025 & $792^3$ \\
huge-1                  & 5.65  & 0.42 & 0.011 & $792^3$ \\
huge-2                  & 5.81  & 0.61 & 0.022 & $792^3$ \\
huge-3                  & 5.66  & 0.82 & 0.042 & $792^3$ \\
huge-4                  & 5.62  & 1.01 & 0.065 & $792^3$ \\
huge-5                  & 5.63  & 1.19 & 0.089 & $792^3$ \\
huge-6                  & 5.70  & 1.38 & 0.12 & $792^3$ \\
huge-7                  & 5.56  & 1.55 & 0.16 & $792^3$ \\
huge-8                  & 5.50  & 1.67 & 0.18 & $792^3$ \\
huge-9                  & 5.39  & 1.71 & 0.20 & $792^3$ \\ 
e5r2                    & 0.13  & 1.57 & 292 & $1200^3$ \\
e5r3                    & 0.61  & 0.52 & 1.45 & $1200^3$ \\
e6r3                    & 5.45  & 0.24 & 0.0039 & $1200^3$ \\
\hline
\multicolumn{5}{l}{\bf Pencil Simulations}\\
ma20                 & 0.91 & 0.20 & 1.56 & $512^3$\\ 
ma23                 & 0.22 & 0.23 & 2.17 & $512^3$\\ 
ma40                 & 0.20 & 0.40 & 8.00 & $512^3$\\ 
\hline
\multicolumn{5}{l}{\bf Pluto Simulations}\\
ma56                 & 0.57 & 0.59 & 2.17 & $512^3$\\ 
ma68                 & 0.19 & 0.68 & 25.51 & $512^3$\\ 
ma75                 & 0.72 & 0.75 & 2.17 & $512^3$\\ 
ma86                 & 0.83 & 0.86 & 2.17 & $512^3$\\ 
\hline \hline
\end{tabular}
\caption{\label{tab:sim}  Description of MHD simulation cubes which some of them have been used in the series of papers the authors have worked on before \citep{2017ApJ...837L..24Y,2017arXiv170303026Y,2018ApJ...865...46L,2020PhRvX..10c1021M,2020NatAs...4.1001Z} the snapshots are taken. }
\end{table}

\section{Application (I): Advancement of the SPA technique}
\label{sec:SPA}

\subsection{Review on the SPA technique}
        In \cite{2020NatAs...4.1001Z} the authors discussed a novel implementation called SPA in obtaining the modes from synchrotron emission maps. Their argue that, since the tensor structures for Alfven or compressible waves being different at the synthesis of the Stokes parameter, the signature of the dominance of the modes are left in the "signature parameter". Let us first recap the formulation of \cite{2020NatAs...4.1001Z} and their main results (See Method section of \citealt{2020NatAs...4.1001Z}). To start with, the authors consider the emissivity of the synchrotron emissions {\it under a locally defined reference frame}:
        \begin{equation}
        \begin{aligned}
            &\epsilon_{xx} = (I+Q)/2 \\
            &= B_{0,\perp}^2\cos^2\phi_s+ 2B_{0,\perp}\cos\phi_sB_i\hat{e}_{xi}+(B_i\hat{e}_{xi})^2
        \end{aligned}
        \label{eq:emis}
        \end{equation}
        where our symbols follow the \cite{2020NatAs...4.1001Z} notations and we employ Einstein notation of summation. Very importantly, the angle $\phi_s$ is the angle between the polarization vector to the currently defined x-axis of the local magnetic reference frame. In \cite{2020NatAs...4.1001Z} they select a small area and compute the change of $\epsilon_{xx}$ as the reference frame rotates. They recognize that the variance of $\epsilon_{xx}$ contains factors that could reflect the relative contributions of MHD modes.

The variance of the emissivity contains the linear term (the first term) and the quadratic term depending on the power of the tensor $\hat{T}$:
\begin{equation}
\begin{aligned}
    &{s}_{xx}=\langle \epsilon_x^2\rangle \propto (2B_{0,\perp}\cos\phi_s)^2 \int d{\bf k}F^2({\bf k})e^{i{\bf k}\cdot{\bf r}} \hat{e}_{xi}\hat{e}_{xi} \hat{T}_{ij}({\bf \hat{k}}) \\ &+ 2 \left(\int d{\bf k} F^2({\bf k})\hat{e}_{xi}\hat{e}_{xi} \hat{T}_{ij}({\bf \hat{k}})\right)^2
\end{aligned}
\end{equation}
\cite{2020NatAs...4.1001Z} pointed out that the linear term, namely the {\bf signature parameter}:
\begin{equation}
    s_{xx}(\phi_s) \propto (2\cos\phi_s)^2\int d{\bf k}F^2({\bf k})e^{i{\bf k}\cdot{\bf r}} \hat{e}_{xi}\hat{e}_{xj} \hat{T}_{ij}({\bf \hat{k}})
\end{equation}
can be expressed in the following format with some constants $a_{xx},b_{xx}$ defined according to MHD theory (See \citealt{2020NatAs...4.1001Z} for details):
\begin{equation}
s_{xx}(\phi_s) = (a_{xx}\sin^2\phi_s+b_{xx})\cos^2\phi_s, \quad \phi_s\in[0,\pi]
\label{eq:zhang_s}
\end{equation}
where the classification parameter $r_{xx}$ is defined as
\begin{equation}
    r_{xx}=\frac{a_{xx}}{b_{xx}}
\end{equation}

In practice, we need to compute the parameter
\begin{equation}
s_{xx, tot} = \frac{Var(\epsilon_{xx})}{4\langle \epsilon_{xx}\rangle}
\label{eq:prac_sxx}
\end{equation}
This term contains both the "linear" term (Eq.\ref{eq:zhang_s}) and the "quadratic" term as defined in \cite{2020NatAs...4.1001Z}.

\begin{figure}
\includegraphics[width=0.49\textwidth]{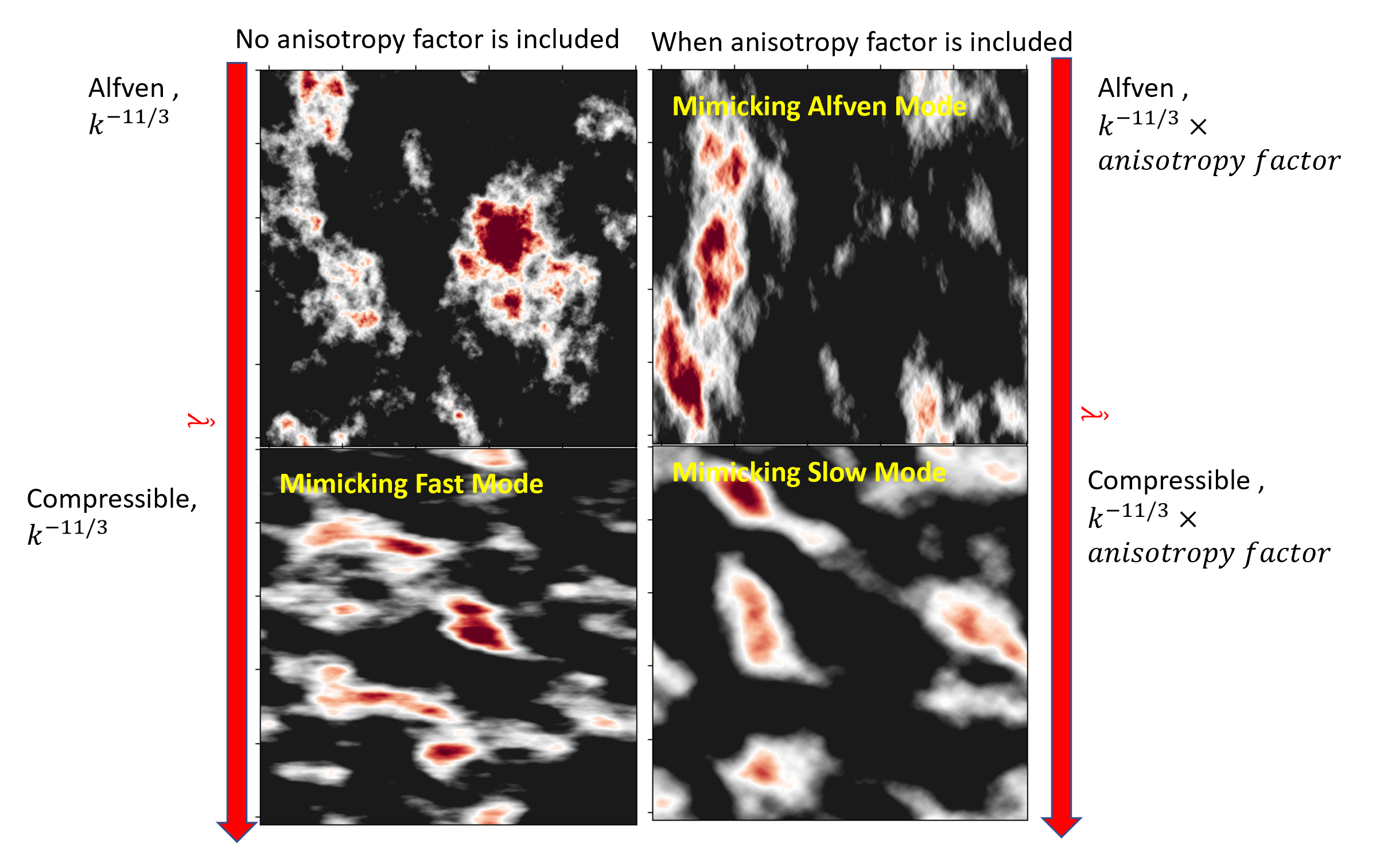}
\caption{An illustration in showing how the observed features of the synchrotron intensities are related to the different weighting of spectrum, anisotropy and frame vectors (See Tab.\ref{tab:theory}) if a strong guided field is given. From the top left: from a $k^{-11/3}$ spectrum plus the Alfven frame vector; top right: a $k^{-11/3}$ spectrum with the Alfven anisotropic factor and also the Alfven frame vector mimicking the Alfven mode; lower left: a $k^{-11/3}$ spectrum with the compressible frame vector, mimicking the Fast mode; lower right: a $k^{-11/3}$ spectrum with the slow mode (See Tab.\ref{tab:theory}) anisotropic factor and also the compressible frame vector, mimicking the slow mode. It is very apparent that both the anisotropy factor and also the frame vector contributes to the observed anisotropy in 2D observables. From synthetic simulations assuming $M_A=0.1$ such that the leakage effect is small. }
\label{fig:mode_projection}
\end{figure}

\subsection{{\hy Alternative ways} to separate the Alfven and compressible modes}
\label{sec:alt}

As discussed in LP12, two-point turbulence statistics contains three types of contributions: The spectrum which measures the cascade as a function of scales, the anisotropy which records whether there is a preferred direction for the cascade to happen, and the tensor structures which records the projection effect of the mode components. From the discussion above, the SPA technique actually did not consider two-point statistics. In particular, Eq.\eqref{eq:prac_sxx} is a one-point statistics, in which the anisotropy of the observable does not play a role in the value of the output. Since all scales are summed up when computing the mean and variances, the spectrum also do not play an important role for Eq.\eqref{eq:prac_sxx}. As a result, a natural guess on how the SPA technique works is the projection effect from the tensor structures. In this scenario using the vector formulation (See Appendix) we can understand quantitatively better how the SPA technique works. Notice that, in the case of one-point statistics, the 2D (i.e. the average operator above) and 3D statistics (which we will consider later below) should be the same.

Let us consider a 3D magnetic field line written as sum of the mean and perturbed contribution in a selected volume $d\Omega$ with $M_{A,d\Omega}\ll 1$:
\begin{equation}
    {\bf H}_i({\bf r}) = \langle {\bf H}_i\rangle + \int d^3 k e^{i{\bf k}\cdot{\bf r}} \sum_{X\in\text{any frame}} C_X(\hat{\bf k},\hat{\bf \lambda}) \hat{\zeta}_X(\hat{\bf k},\hat{\bf \lambda})
\end{equation}
From now on we are going to choose the frame to be the PCA frame (See Appendix \ref{subsec:T}, Fig.\ref{fig:illus}), and assuming the line of sight direction is at the z-axis and {the magnetic field in the plane of sky defines the x-axis}. The x-component magnetic field dispersion, which is just the mean value of the emissivity subtracted by a constant (cf.Eq.\eqref{eq:emis}), is given by:
\begin{equation}
    \langle \delta H_x^2\rangle = 2\pi\int d^3 k \sum_{X\in C,A} C_X^2(\hat{\zeta}_X\cdot \hat{x})^2
    \label{eq:dH}
\end{equation}
Notice that the only difference between the compressible and the Alfven component can be observed when we expand the dot product for the above equation:
\begin{equation}
    \begin{aligned}
    \hat{\zeta}_A \cdot \hat{x} &= \frac{(\hat{\lambda}\cdot \hat{z})(\hat{k}\cdot \hat{y})}{ |\hat{k}\times\hat{\lambda}|}\\
    \hat{\zeta}_C \cdot \hat{x} &= -(\hat{\lambda}\cdot \hat{x}) \frac{1-(\hat{k}\cdot\hat{x})^2}{ |\hat{k}\times\hat{\lambda}|}
    \end{aligned}
    \label{eq:ACdot}
\end{equation}

We can model Eq.\eqref{eq:dH} via the frame definition of $\phi_s$ in Eq.\eqref{eq:emis}, where the frame angle $\phi_s=0$ when the projection of magnetic field is along the x-axis:
\begin{equation}
    \langle \delta H_x^2\rangle = A_{xx} \cos^2\gamma + B_{xx} \sin^2\gamma \cos^2\phi_s
    \label{eq:SPA_alf}
\end{equation}
where $\cos\gamma = \hat{\lambda}\cdot \hat{z}$ is the line of sight angle, and $A_{xx},B_{xx}$ are:
\begin{equation}
    \begin{aligned}
    A_{xx} &=  2\pi \int d^3 k C_{A,obs}^2 \left(\frac{(\hat{k}\cdot \hat{y})}{ |\hat{k}\times\hat{\lambda}|}\right)^2\\
    B_{xx} &= 2\pi \int d^3 k C_{C,obs}^2 \left(\frac{1-(\hat{k}\cdot\hat{x})^2}{ |\hat{k}\times\hat{\lambda}|}\right)^2
    \end{aligned}
    \label{eq:AB_general}
\end{equation}
The factors within the bracket of each equation above are the {\it geometric factors} as discussed in LP12. Here we consider the general case of the leakage, which applies to both Alfven and compressible modes (See \S\ref{subsec:leakage}), i.e. the observed amplitudes of Alfven and compressible modes $C_{A,obs},C_{C,obs}$ undergo an orthogonal rotation of angle $M_A<1$ (See Eq.\ref{eq:postulate}):
\begin{equation}
    \begin{aligned}
    C_{A,obs} &\approx C_A \cos M_A - C_C \sin M_A\\
    C_{C,obs} &\approx C_A \sin M_A + C_C \cos M_A
    \end{aligned}
\end{equation} The expressions inside the brackets of $A_{xx},B_{xx}$ are the geometric factors that considered in both LP12 and \cite{2020NatAs...4.1001Z}. We can see from Eq.\eqref{eq:SPA_alf} that the contributions of Alfven and compressible modes are separated when one considers the frame rotation even for $\langle \epsilon\rangle$. It is not necessary to compute Eq.\eqref{eq:prac_sxx} in extracting the contributions of the modes. Moreover, we can now quantify the contributions of modes via Eq.\eqref{eq:AB_general} by using the modes for $C_{A,C}$ by simply comparing the values of $A_{xx}$ and $B_{xx}$ while analyzing the observed synchrotron emission. In particular, if $M_A$ is small and there is no compressible mode, then $B_{xx}=0$. i.e. that contribution of the Alfven mode to $\langle \epsilon \rangle$ is frame independent (i.e. rotating the x-y plane does not alter the result) since $(\hat{\lambda}\cdot \hat{z})$ cannot be changed due to frame rotation, while that for compressible mode is a frame dependent quantity since $\hat{\lambda}\cdot \hat{x}$ is a function of the reference frame. 

\subsection{The SPA technique for synchrotron emissions with significant Faraday Rotation}
\label{sec:SPA_FR}

{ In \cite{2020NatAs...4.1001Z} they study the effects of Faraday rotation to the SPA technique. Pictorially the Faraday depolarization effects shields information up to a certain distance along the line of sight. This distance has been adequately discussed in \cite{LP12} in the presence of galactic MHD turbulence and is called the Faraday screening effect \citep{LY18b}. Qualitatively, the SPA technique can only determine the mode fraction before the Faraday screen. However, we would like to perform the analysis based on the formalism of \cite{LY18b}. 

}

In general, the synchrotron emission depends both on the distribution of relativistic electrons 
\begin{equation}
N_e({\cal E})d{\cal E}\sim {\cal E}^{\alpha} d{\cal E},
\end{equation}
 with intensity of the synchrotron emission being
\begin{equation}
I_{sync}({\bf X}) \propto \int dz B_{\perp}^\gamma({\bf x})
\end{equation}
where ${\bf X} = (x,y)$ is the 2D position of sky (POS) vector and $B_{\perp} = \sqrt{B_x^2 + B_y^2} $ being the magnitude of the magnetic field perpendicular to the LOS $z$. In general, $\gamma=0.5(\alpha+1)$ is
a fractional power, which was a serious problem that was successfully addressed in LP12. LP12 proves that the statistics of $I(\alpha)$ is similar to that of $I(\alpha=3)$. Therefore it suffices to discuss the statistical properties of the case $\alpha=3$.

Per \cite{LP12}, Synchrotron complex polarization function {\it with Faraday rotation} is given by:
\begin{equation}
    P_{synch}({\bf R}) = \int dz \epsilon_{synch} \rho_{rel}B^2e^{2i\left(\theta({\bf R},z)+C\lambda^2\Phi(R,z)\right)}
\end{equation}
where $\epsilon_{synch}$ is the emissivity of synchrotron radiation,
\begin{equation}
    \Phi(R,z) = \int_\infty^z dz' (4\pi)^{-1/2}\rho_{thermal}({\bf R},z) B_z({\bf R},z) {\rm rad~m^{-2}}
    \label{eq:chap1.frm}
\end{equation} is the Faraday Rotation Measure \footnote{It is usually more convenient to use $H_z=B_z/\sqrt{4\pi}$ for analysis. }. Notice that $\rho_{rel}$ is the relativistic electron density, while $\rho_{thermal}$ is the thermal electron density. The C-factor $\approx 0.81$  (Lee et.al 2016). The projected magnetic field orientation is then given by:
\begin{equation}
    \theta_B = \frac{\pi}{2} + \frac{1}{2} \tan^{-1}_2(\frac{U}{Q})
    \label{eq:chap1.Bangle}
\end{equation}
where $\tan^{-1}_2$ is the 2-argument arc-tangent function.

For frequencies lower than $O(1GHz)$, the amplitude of the Faraday Rotation measure will exceed $2\pi$. The physical picture of the synchrotron polarization with Faraday rotation measure can be understood as: photons that passes through a section of ISM has to experience a certain amount of phase shift. If this phase shift exceeds $2\pi$, all information coming from the source is completely lost. Therefore an important concept called the {\bf Faraday screening} emerges, which indicates the maximal line of sight distance that the observed synchrotron emissions can measure in the presence of line of sight magnetic field. In the case of sub-Alfvenic turbulence, the source term $P_i = \rho_{rel}\exp(2i\theta({\bf R},z))$ is dominated by the mean field rather than the fluctuating one. The two regimes: (1) strong and (2) weak Faraday Rotation depend on whether the ratio $L_{eff}/L$,  is smaller (strong) or larger (weak) than unity:
\begin{equation}
\label{eq:el}
\frac{L_{eff}}{L} \sim \frac{1}{\lambda^2L} \frac{1}{\phi}
\end{equation}
where $\phi= \max(\sqrt{2} \sigma_\phi,\bar{\Phi})$ with $\sigma_\phi$ is the dispersion of random magnetic field. The difference between the two regimes are, the Faraday rotation and the emission happens together in the former regime ($\phi= \sqrt{2} \sigma_\phi$), while the latter has the Faraday rotation happens after the emission of the polarization. We shall name the two regimes "Variance-driven Faraday Rotation" (VFR) and "Mean-field Faraday Rotation" (MFR), respectively. Notice that both regimes have been considered in \cite{2020NatAs...4.1001Z}.

\begin{figure}
\includegraphics[width=0.48\textwidth]{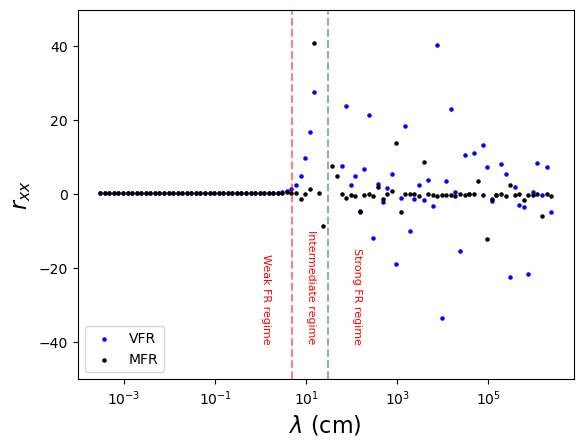}
\caption{A figure showing how the values of $r_{xx}$ varies as a function of $\lambda$ in the presence of Faraday Rotation $\propto \lambda^2 \int dz \rho B_z$ for both Variance-driven Faraday Rotation (VFR) and mean-field driven Faraday Rotation (MFR).  }
\label{fig:spg}
\end{figure}

Fig.\ref{fig:spg} shows a plot on how VFR and MFR could change the value of $r_{xx}$. For this current plot we {\it intentionally} plot $r_{xx}$ with values that are not typically considered in previous literature (See, e.g. \citealt{2020NatAs...4.1001Z}, $r_{xx}\in [-1,1]$). This allows us to better characterize whether the value of $r_{xx}$ came from the effect of compressibility or from Faraday rotations. We can observe from Fig.\ref{fig:spg} that there are two new regimes of $\lambda$ that could make $r_{xx}$ fluctuates well beyond the values previously considered in \cite{2020NatAs...4.1001Z}. From Fig.\ref{fig:spg} we classify the ranges of values of $\lambda$ via the fluctuations of $r_{xx}$ into three regimes: he "weak" regime correspond to the case where $r_{xx}$ is small ($\in [-1,1]$ as in \citealt{2020NatAs...4.1001Z}). The intermediate regime correspond to the case where $r_{xx}$ start to grow exponentially, and the strong regime correspond to the case where the $r_{xx}$ basically loses traces on the compressibility. We can see that obviously the technique of SPA does not work when we are in the strong regime. However an interesting question is whether the SPA technique actually works in the intermediate regime { which will be the subject for future studies}.

\section{Application (II): A self-consistent line of sight angle tracing method via structure functions of $I+Q$ and $I-Q$}
\label{sec:gamma}

The second application that we will deliver in this paper would be the retrieval of the mean global line of sight angle $\gamma$. In the case of synchrotron/dust polarization, we have adequate information to estimate $\gamma$ by considering the structure functions of both $I+Q \propto \int dz B_x^2$ and $I-Q\propto \int dz B_y^2$. For the following subsections, we will assume that the global mean field within the sampling area is $\parallel$ to x-axis. One could always rotate the frame in Stokes parameter space to have the above condition satisfied. 

\subsection{Why $\gamma$ is encoded in the statistics of $I+Q$ and $I-Q$?}

\begin{figure}
\centering
\includegraphics[width=0.49\textwidth]{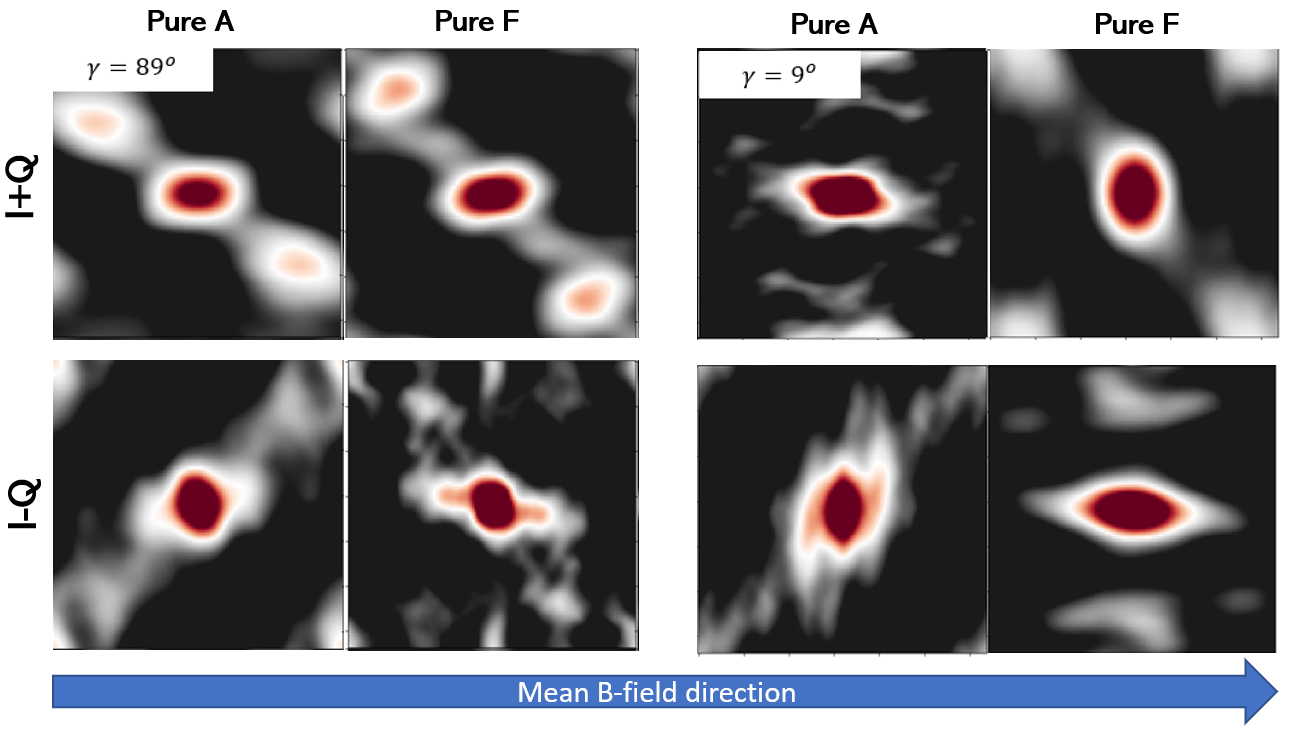}
\caption{\label{fig:multipole_illus}  A set of figures showing how the orientation of anisotropy for $I+Q$ and $I-Q$ is related to the line of sight angle $\gamma$ for pure A (Alfven) and F (compressible, see \citealt{LP12}) type tensor. The key difference between the case of $\gamma\rightarrow \pi/2$ (correspond to the case when $B\perp$ LOS) and $\gamma\rightarrow 0$ is that, the anisotropies of $I+Q$ and $I-Q$ for pure A and F tensors are similar for the former case, while for the latter case the anisotropies of pure A and F tensors are exactly opposite. }
\end{figure}

\begin{figure*}
\centering
\includegraphics[width=0.9\textwidth]{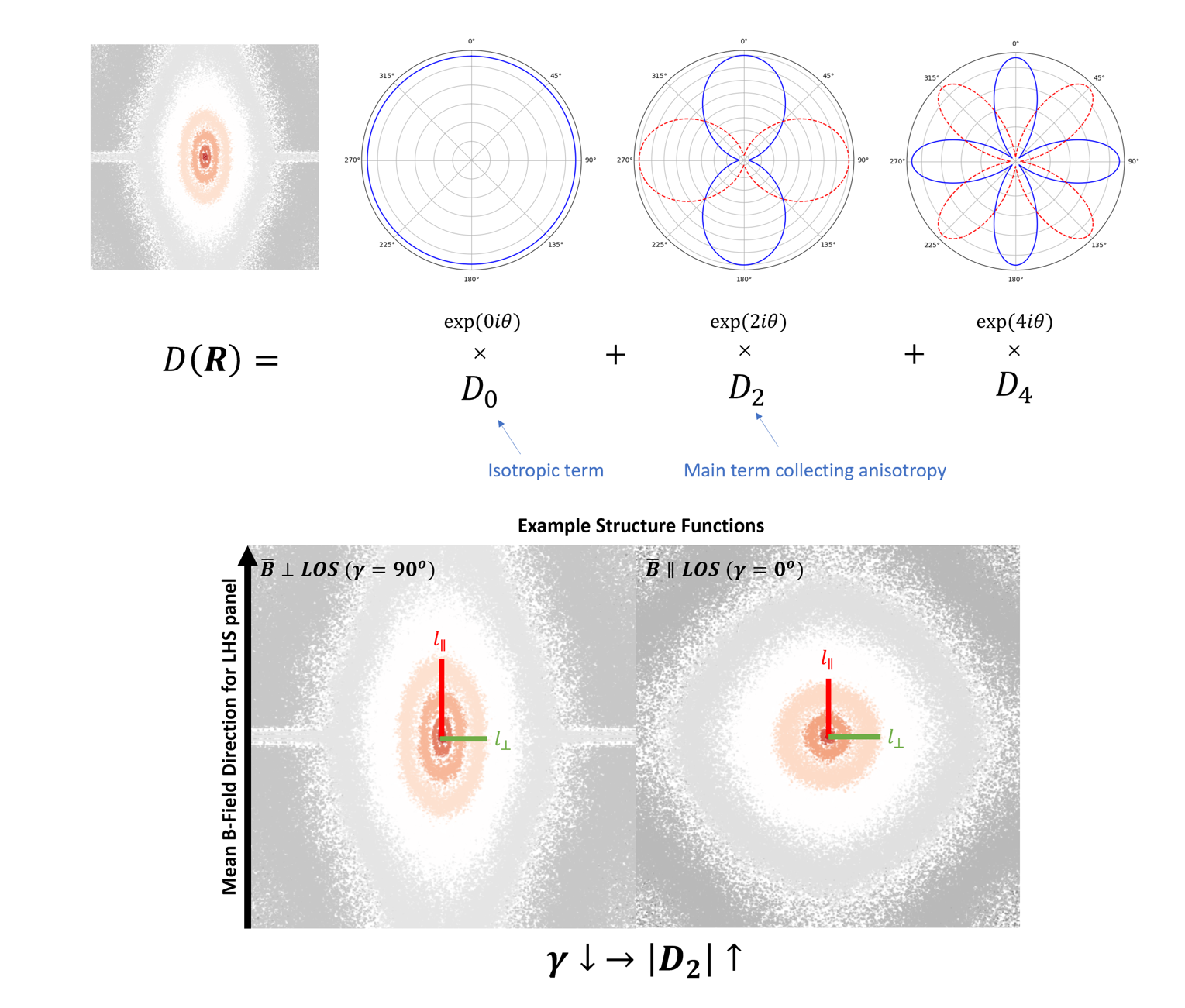}
\caption{\label{fig:multipole} A set of visualizations showing how the structure function of a certain variable $D({\bf R})$ can be visually decomposed as the linear combination of the multipole moments $D_n$, and how the multipole moments should be physically correlated to the relative angle between the line of sight and mean magnetic field  $\gamma$. The multipole moments collects the relative weight on the shapes that are specifically defined with the angular function $\exp(in\theta)$. In particular, $D_0$ records the weights of the isotropic components of the structure functions, while $D_2$ records the first order directionless anisotropy. Since empirically structure functions are mostly elliptical-like, $|D_2|$ must be non zero. modern turbulence theory predicts that the observed anisotropy would be a function of $\gamma$. When $\bar{B} \parallel$ LOS, then the structure function should be isotropic. While $\bar{B}\perp $LOS, the structure function should be anisotropic. Therefore under the framework of multipole moments, the absolute amplitude of $D_2$ should be a function of $\gamma$.  }
\end{figure*}

The essence of on why $\gamma$ is encoded in $I+Q$ and $I-Q$ is based on the fact that tensor formulation (c.f. Eq.\ref{eq:2}) contains different expressions for observables that $\parallel B$ and $\perp B$. In Fig. \ref{fig:multipole_illus}, we present a set of figures showing the anisotropy of $I+Q$ and $I-Q$ for both A and F type contributions (c.f. Fig.\ref{fig:illus}). We present two extreme cases for $\gamma$ in Fig.\ref{fig:multipole_illus} \footnote{Notice that the projection of pure Alfven wave fluctuations when $\gamma$ is exactly $90^o$ will vanish, see \cite{ch5} for the analysis.} that is sufficient to illustrate the differences of behaviors for the anisotropy of A and F type fluctuations. The left group of figures in Fig.\ref{fig:multipole_illus} shows the case when $\gamma = 89^o$, while the group of figures on the right shows the case when $\gamma = 9^o$. We can observe from Fig.\ref{fig:multipole_illus} a few interesting phenomena which is not covered in previous anisotropy studies:
\begin{enumerate}
\item The anisotropies of A and F type tensor do not necessarily align with the mean magnetic field direction. We discussed this effect already from Fig.\ref{fig:mode_projection}. The reason behind is that both the anisotropy and tensor contribution are anisotropic (c.f. \S \ref{sec:alt}). However, the direction of anisotropy for the tensor contribution (with the Alfven leakage in effect) does not necessary be parallel to B-field and is a function of $\gamma$. Notice that the change of anisotropy is highly tied with the $\gamma$ value (See Fig.\ref{fig:multipole})
\item For the case of pure Alfven fluctuations, the anisotropy is more or less parallel to magnetic field for $I+Q$, while $\perp B$ for $I-Q$. Yet, the compressible mode does not carry the same trend as its Alfven counterpart: When $\gamma \approx 9^o$, the F-type anisotropy for $I+Q$ is actually $\perp B$, while that for $I-Q$ is $\parallel B$. In contrast, when $\gamma \approx 89^o$ the F-type anisotropy varies very similarly to that of the Alfven counterpart. 
\item The measurement of relative anisotropies between $I+Q$ and $I-Q$ allows us to characterize the $\gamma$ value. From Fig.\ref{fig:multipole_illus} we can see that if we consider the anisotropies of $I+Q$ and $I-Q$ at $\gamma \approx 89^o$, $I+Q$ tends to be parallel to magnetic field, while that for $I-Q$ tends to be perpendicular to magnetic field. We utilize the formulation in Appendix \ref{ap:anisotropy} that the minor-to-major axis ratio $l_\perp/l_\parallel=\sqrt{1-\epsilon^2}$, which the eccentricity $\epsilon$ is related to the quadropole-to-monopole ratio $|D_2/D_0|$ via Eq.\ref{eq:d2d0_to_e}. The quadropole-to-monopole ratio is the key parameter in parametrizing the anisotropy in previous literature \citep{LP12,LP16,KLP16,KLP17,ch5}.
\end{enumerate}

\begin{figure}
\centering
\includegraphics[width=0.49\textwidth]{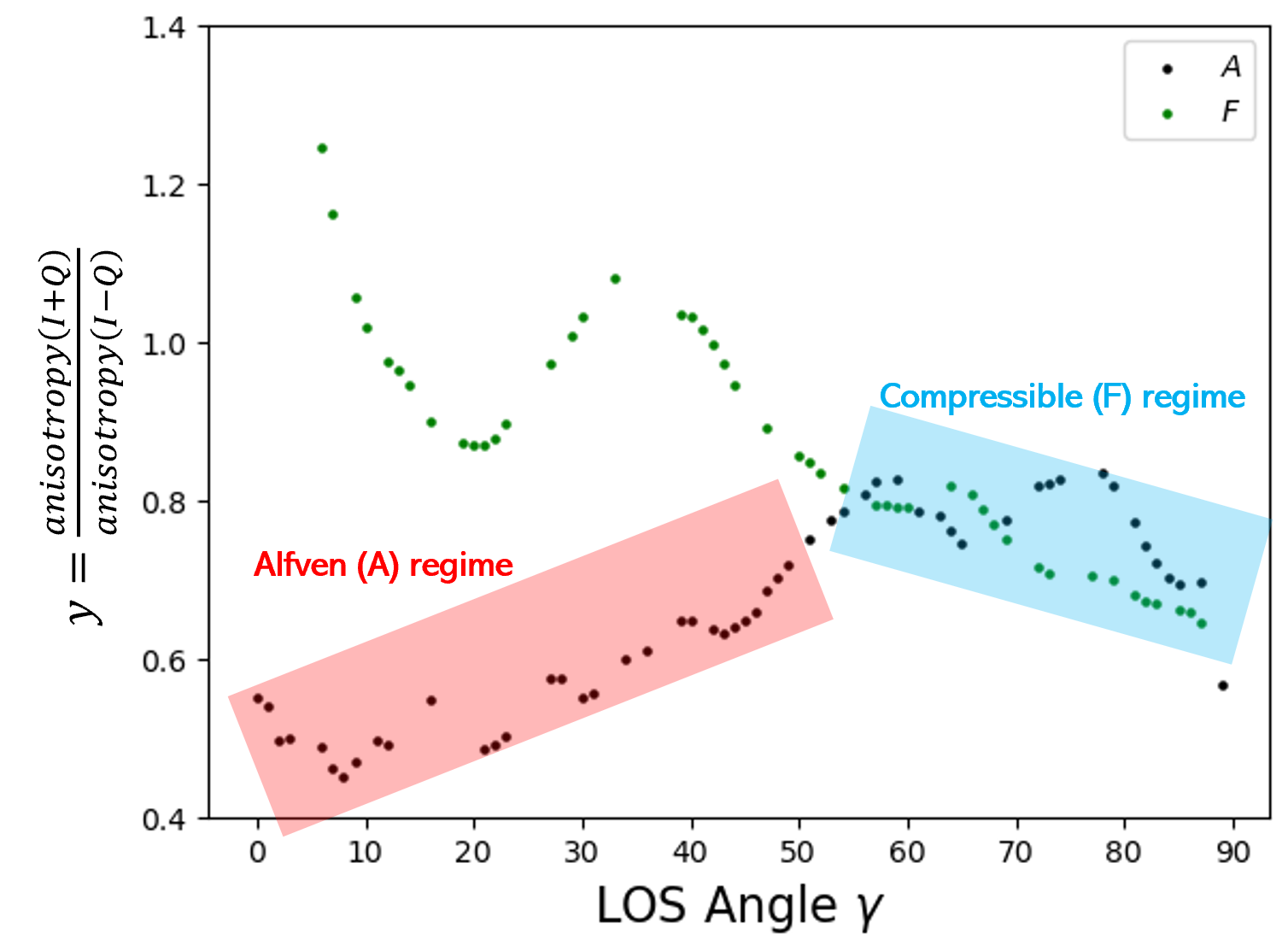}
\caption{\label{fig:y_model} A figure showing the characterization of the relative anisotropy index ($y = \text{anisotropy}(I+Q)/\text{anisotropy}(I-Q)$) as a function of the line of sight angle $\gamma$. As we outlined in Fig.\ref{fig:multipole_illus}, the relative anisotropies for A and F type fluctuations are different when $\gamma$ is different. For the case of $\gamma \rightarrow 90$, we expect the anisotropies of A and F type tensor fluctuate in the same way, which is illustrated as the light blue box in the figure. However, when we are looking at small $\gamma$ limit, the anisotropies of A and F type tensor went completely opposite, which is highlighted by the red box in the figure. We denote these two regimes the "compressible" and the "Alfven" regime respectively. from numerical simulation "e5r2" (See Tab.\ref{tab:sim}.) }
\end{figure}

\begin{figure*}
\centering
\includegraphics[width=0.9\textwidth]{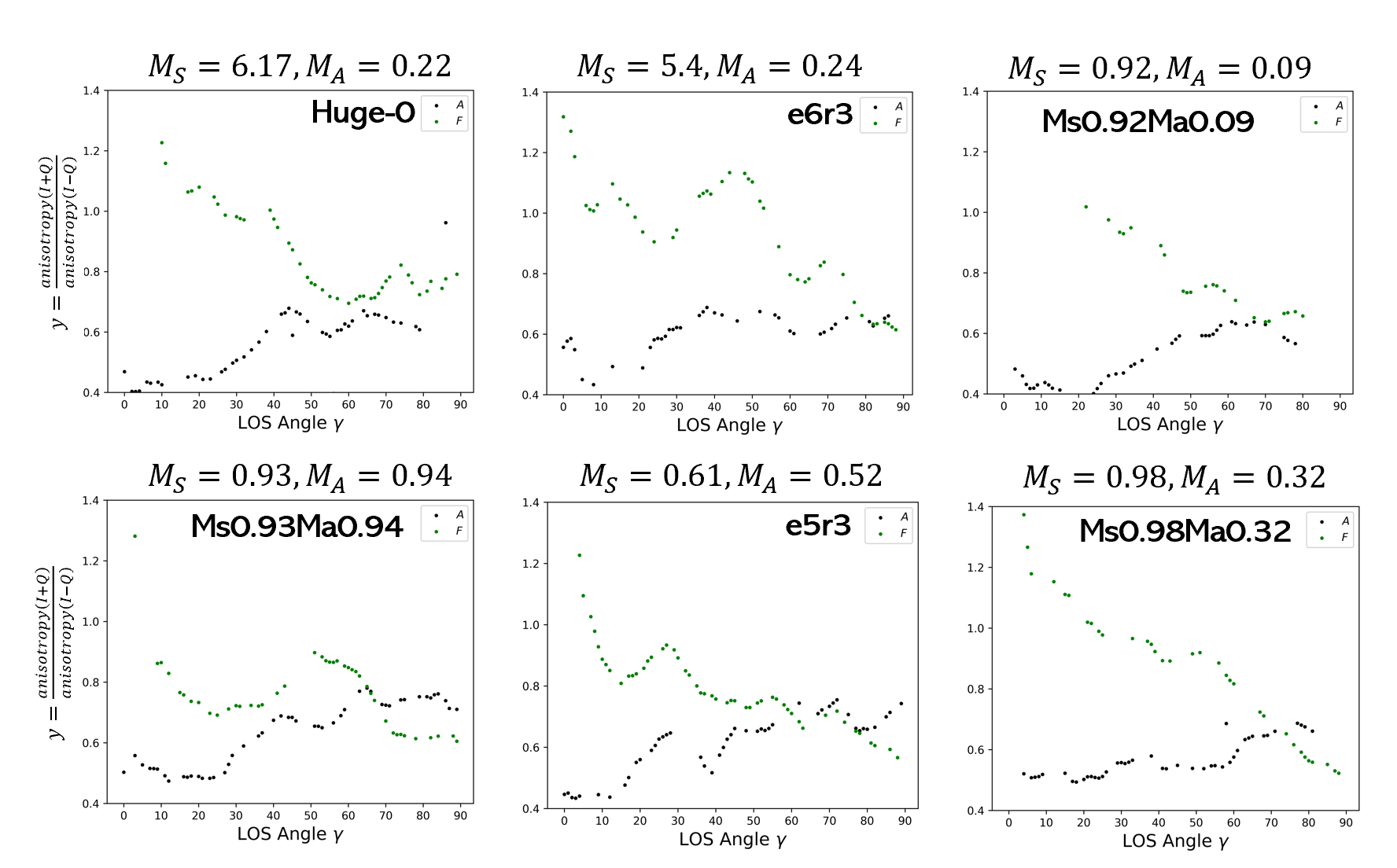}
\caption{\label{fig:y_result}  A set of figures showing the universality of our finding ($y$ as function of the line of sight angle $\gamma$ in degrees) in Fig.\ref{fig:y_model} in 6 numerical simulations from Tab.\ref{tab:sim}, which covers a large range of value of $M_{s,A}$.   }
\end{figure*}

\subsection{Formalism via \S \ref{subsec:leakage}}

We will start from the parameters $I+Q$ and $I-Q$ in which we will assume the projected mean field is right now along the $x$ direction \footnote{For a general magnetic field configuration, one could always consider the combination $I+(Q\cos(2\phi_{pol})-U\sin(2\phi_{pol}))$, where we perform an inverse orthogonal transform with twice of the polarization angle $2\phi_{pol} = \tan^{-1}_2{U/Q}$ for this analysis.}. For the case of $I+Q$, we adopt the structure function expression from Eq.(E20) of \cite{ch5}:

\begin{equation}
\begin{aligned}
D_{I+Q}({\bf R}) &= \langle (B_x({\bf R}+{\bf R}')-B_x({\bf R}'))^2\rangle_{{\bf R}'}\\
&= \frac{1}{2\pi^2} \! \int \!\! d^2 K \left( 1 - e^{i \mathbf{K} \cdot \mathbf{R}}\right) \times\\
&\Big[ 
A(K,\sin\gamma \cos\phi_K) \frac{ {\color{red} \cos^2\gamma} \sin^2\phi_K}{1 - \sin^2\gamma \cos^2\phi_K}  +  \\ &F(K,\sin\gamma \cos\phi_K)
\frac{{\color{red} \sin^2\gamma} \sin^4\phi_K }{1 - \sin^2\gamma \cos^2\phi_K} \Big]
\end{aligned}
\end{equation}

where those factors are simply the expressions of $\zeta_A\zeta_A$ and $\zeta_F\zeta_F$ in the global frame of reference (i.e. after leakage). The main takeaway here is, This D factor depends on the following form

\begin{equation}
D_{I+Q}({\bf R}) \sim \bar{A}({\bf R})\cos^2\gamma + \bar{F}({\bf R})\sin^2\gamma
\label{eq:dipq}
\end{equation}

Similarly, the structure function for $I-Q$ can be also modelled similarly as:
\begin{equation}
D_{I-Q}({\bf R}) \sim \bar{A}({\bf R})\sin^2\gamma + \bar{F}({\bf R})\cos^2\gamma
\label{eq:dimq}
\end{equation}

Based on Fig.\ref{fig:multipole_illus} we can see that the construction:
\begin{equation}
\bar{y} = \frac{\text{Anisotropy}(D_{I+Q})}{\text{Anisotropy}(D_{I-Q})}
\label{eq:y_central}
\end{equation}
contains the information on $\gamma$. Here we take the convention that $\text{Anisotropy}(D)>1$ when the anisotropy of structure function is parallel to the global magnetic field direction, and vice versa. In particular, from Fig.\ref{fig:multipole_illus} we expect that $\bar{y}_A>1$ for all $\gamma$, while that for F-type contribution changes from smaller than 1 to greater than 1. Detecting the value of $\bar{y}$ for compressible modes (in global frame of reference) detected in observation is the key to extract the value of $\gamma$.

The key reason why we consider the ratio of structure functions instead of individual quantity is because, from our expressions {\it in the global frame of reference}, the structure function of some observables carries factors on spectrum, anisotropy and tensors. For the case of structure functions of $I+Q$ and $I-Q$, their only difference is coming from the tensor factor as spectrum and anisotropy factors are fixed once the turbulence is set-up.

To proceed with our analysis, we consider the multipole expansion up to quadrupole (See Appendix \S \ref{ap:anisotropy} for the condition for the expansion. In particularly, the expansion is valid only for $M_A\sim 0.5-1.0$.). Formally we can express the anisotropy function that we defined above with the monopole and quadrupole coefficients $D_0,|D_2|$:
\begin{equation}
    \text{Anisotropy}_{M_A \in [0.5,1]}\approx \text{sign}(\text{Anisotropy})\times \frac{D_0-|D_2|}{D_0+|D_2|}
\end{equation}
Recall from the previous discussion that the factors $d_{0,2}$ can be literally written as the spectrum, anisotropy and the tensor contribution, and the first two contributions are cancelling out under our treatment, we can formally write $y$, which is the quadrupole approximation of $\bar{y}$ to be (c.f. Eq.(E30) of \citealt{ch5}):
\begin{equation}
\begin{aligned}
y &= \frac{\text{Anisotropy}(I+Q)}{\text{Anisotropy}(I-Q)}\\
&=\left(\frac{D_0-|D_2|}{D_0+|D_2|}\right)_{I+Q} \left(\frac{D_0-|D_2|}{D_0+|D_2|}\right)_{I-Q}^{-1}
\end{aligned}
\end{equation}
where we notice that under our current configurations, $I+Q \sim {\cal L}_z b_x^2$ and  $I-Q \sim {\cal L}_z b_y^2$ where $L_z$ is the length of the integration. Keeping only the tensor term, we will have an expression that is purely based on $W_{I,L}$ in Eq.\ref{eq:our_WIWL}, and also functions of $\gamma$ (See Eqs.\ref{eq:dipq} and \ref{eq:dimq}).

Fig.\ref{fig:y_model} shows how numerically the factor $y$ depends on the line of sight angle $\gamma$ for Alfven mode (black) and the compressible mode (green). We notice that the qualitative phenomenon happened in Fig.\ref{fig:multipole_illus} is exactly described by $y$ for compressible modes: $y<1$ for $\gamma\rightarrow 90^o$, while $y>1$ for $\gamma\rightarrow 0^o$. We recognize that there are fluctuations in terms of the variation of $y$ relative to $\gamma$ for the compressible case. Surprisingly, the Alfven mode $y$ also exhibits some interesting properties that we can exploit in obtaining $\gamma$ in observation. Notice that $y$ for Alfven mode stays $<1$ from what we observe in Fig.\ref{fig:multipole_illus}, we see that the Alfven mode's $y$ has very similar trend when $\gamma\gtrapprox 55^o$ , but when $\gamma \lessapprox 55^o$, the Alfven mode $y$-value went exactly opposite to that of compressible mode. Moreover, we observe that the change of values of $y$ as a function of $\gamma$ is more or less monotonic if we consider $\gamma \lessapprox 55^o$ and $\gamma \gtrapprox 55^o$. Notice that the modes that we are talking about here are all in the global frame of reference.

To see whether the trend that we observed in Fig.\ref{fig:y_model} is robust, we select some of the numerical cubes from Tab.\ref{tab:sim} and to plot $y$ as a function of $\gamma$ for both $A$ and $F$ type contribution and plot it as Fig.\ref{fig:y_result}. The selected numerical cubes cover a wide range of sonic and Alfvenic Mach numbers. We can see from Fig.\ref{fig:y_result} that the trends of the two curves are very similar to that of Fig.\ref{fig:y_model}. Furthermore, the exact values of $y$ are also very similar across different turbulent conditions. Originally, the formalism of $A$ and $F$ type tensor applies only for $M_{s,A}<1$. However, we perform the calculation of $y$ also for supersonic sub-Alfvenic simulations, which is closer to the environment of molecular clouds (See, e.g. \citealt{2011piim.book.....D}) and still observe the same trend. We therefore conclude that the $\bar{y}$ parameter tracers $\gamma$. In fact, we observe from Fig.\ref{fig:y_result} that when the plasma $\beta\propto M_A^2/M_s^2$ is smaller, it is easier to recover the trend that we see in Fig.\ref{fig:y_model}.

At last, we provide the empirical formula (units in degrees) for the case of low $\beta$ ($\beta<1$). For $\gamma < 40^o$
\begin{equation}
\begin{aligned}
y(F) &\sim 1.2 - \gamma/40\times0.4\\
y(A) &\sim 0.4 + \gamma/40\times0.2
\end{aligned}
\end{equation}
for $\gamma > 40$ degrees
\begin{equation}
\begin{aligned}
y(F) &\sim 0.8 - (\gamma-40)/50\times0.2\\
y(A) &\sim 0.8
\end{aligned}
\end{equation}
The full study on how the y-parameter can be applied to situation with different mixture of driving will be discussed in Malik et al. (in prep).

\section{Discussion}
\label{sec:discussion}
\subsection{The importance of Alfven leakage for mode decomposition}

{\kh The analysis of} turbulence properties generally from observations requires the consideration of the local-to-global frame problem, which is modelled as the "magnetic field wandering problem". While the theory of MHD turbulence is well-established, how the local scaling laws are projected globally is still mysterious, despite models have been proposed from both \cite{LP12} and \cite{2020arXiv200207996L}. Here, we propose the first physical model in explaining how the wandering of magnetic field happens {\kh when projected along the line of sight}, and how we could utilize the magnetic field wandering in deducing a number of important physical quantities such as the line of sight angle and also the mode fractions.

The problem of the local-to-global frame transition in theoretical MHD turbulence studies have puzzled the community for a while. While the anisotropic scaling $k_{\parallel}\sim k_{\perp}^{2/3}$ is well motivated from the simple constant energy cascade and critical balance condition (GS95), we cannot retrieve the local scaling from the global frame of reference. In fact, the global correlation function usually gives a constant scaling rather than a geometrically-driven, size-dependent scaling as predicted by GS95. Before the availability of MHD simulations (e.g. \citealt{2003MNRAS.345..325C,2005ApJ...624L..93B}), it is not yet possible to validate the GS95 relation even from numerical simulations. 

The more puzzling effect comes when $M_A$ is very large. Traditionally the numerical test on GS95 are done in small $M_A$ systems and in small scales. However as we see from the previous sections, in moderate and small $k$ the Alfven mode acquired from the \cite{2003MNRAS.345..325C} decomposition method contains non-negligible contributions along the $\hat{\zeta}_C$ vector, indicating {\it the presence of anomalous compressive wavevector even after Alfven mode decomposition.} The only plausible reason why this happens is because the mode decomposition method from \cite{2003MNRAS.345..325C} is done on a global frame of reference. As a result when we are looking at small scales, {\it in average} the mean field is not very different from its local field. Yet for larger scales the mean field is very different from the local field, so that there exists anomalous compressible terms even the data are supposed to be "Alfven modes" according to \cite{2003MNRAS.345..325C}.  We named this effect "Alfven leakage" in our previous section since this effect happens even for Alfven waves as long as the Alfvenic Mach number is not zero.

In this paper, we further show that the Alfven leakage effect is a global function of $M_A$. In fact, the presence of the leakage effect suggests that the mode decomposition method by \cite{2003MNRAS.345..325C} should subject to the a correction term for moderate and small $k$. However since most of the calculation from \cite{2003MNRAS.345..325C} are done in small scales, i.e. large $k$, the results of their work are not affected.

\subsection{The {\hy importance of tensor forms} to the SPA technique and general turbulence studies}


The novel invention of the SPA technique (\cite{2020NatAs...4.1001Z}) utilizes the fact that the tensor projections have different contributions for Alfven and compressible modes to identify them in observations. This work further strengthens their argument through the use of Alfven leakage picture and suggests a few important improvements to their method. For instance, it is not necessary to compute the parameter $s_{xx}$ as in Eq.\ref{eq:prac_sxx} to distinguish the modes. The tensor properties are encoded in the Stokes parameters and thus ignoring the tensor contribution would make dramatically different predictions in astrophysical applications. 

One very important factor that is accounted by \cite{2020NatAs...4.1001Z} is the use of one point statistics under {\kh Stokes frame transformation}. The traditional turbulence statistical studies usually utilize multi-point statistics since they are either directly related to the spectra (e.g. two-point) or is used to validate scaling relations for higher order structure functions (e.g. Kolmogorov 4/5 law). The reason of why single point statistics was not useful before is because the spectrum and anisotropy are the main characteristics of turbulence studies for the past 60 years. However, how the tensor projection affects the geometry of the structures for each of the turbulence variable is not really explored. {\hy Tensor forms of turbulence modes were not much explored beyond the physics of cosmic rays\citep{2002cra..book.....S,2002PhRvL..89B1102Y,2004ApJ...614..757Y}}. In fact, the previous anisotropy analysis also did not consider what is the statistics of a single component of a 3D turbulence, i.e.tensor projection, after projection along the line of sight. While the series of papers by \citeauthor{LP12} started to consider how the single component statistics works, not until recently did both numerically (e.g. \citealt{2018ApJ...865...46L}) and observationally \citep{2020NatAs...4.1001Z} found the effect of tensors to be that important during single component projection. The anisotropy of projected fast modes with the direction opposite to the Alfvenic anisotropy was shown in \cite{LP12}. In fact, one of the most common belief that is circulating in the earlier studies of MHD turbulence theory (e.g. the discussion section of \cite{2017ApJ...842...30L}) is the presumption that the projection of the observables (e.g. velocities, magnetic field) from fast modes will be isotropic since the fast modes in 3D are. This is empirically proven wrong by \cite{2018ApJ...865...46L} through the development of velocity gradient and also utilized through the development of SPA in \cite{2020NatAs...4.1001Z}. 

In fact, in a number of astrophysical applications, the observables are constructed through not all three directions of velocities or magnetic field, but just some of them. For instance, the Davis-Chandrasekhar-Fermi (DCF) technique utilizes both the line of sight velocity dispersion and the plane of sky polarization angle dispersion to estimate the magnetic field strength through the use of Alfven relation (See also \citealt{2016ApJ...821...21C}). However, as found in \cite{ch5}, the direction of the velocity and magnetic field fluctuations as collected in DCF technique are exactly perpendicular to each other. Moreover, in this work we also show both analytically and numerically that the tensor term contains anisotropy and can be dominant as long as the $\gamma$ fulfills some conditions. As a result, one should not ignore the contributions of the tensor term in studying the properties of MHD turbulence.

\subsection{The use of high pass filter?}

In MHD turbulence studies, there are a few length scales that determine whether the underlying turbulence is hydrodynamic or GS95-like. It is a general phenomenon that for 3D saturated turbulence the small scale fluctuations are GS95 like. However for both sub-Alfvenic and super-Alfvenic there exist a transition scale that the turbulence becomes non-GS95 like. For instance, for the sub-Alfvenic case there exist the transition from weak to strong turbulence (\citealt{2003MNRAS.345..325C,2020PhRvX..10c1021M}) at the length scale $LM_A^2$, while for the super-Alfvenic case above the scale $LM_A^{-3}$ the turbulence is hydrodynamic. This might suggest that the removal of large scale fluctuations could allow observers to obtain the desired GS95 statistics with the use of high pass filters. 

{\kh
However upon projection the high pass filter in 2D acts a little bit differently compared to 3D. Fundamentally the high pass filter (HPF) in 3D serves as the high frequency extractor.  As noticed in \cite{2020arXiv200207996L} , HPF in 2D acts as a lower bound of the HPF in 3D, i.e. if we explicitly want $ K = \sqrt{k_x^2+k_y^2} > K_0$, this will automatically apply to $k = \sqrt{k_x^2+k_y^2+k_z^2} \ge K > K_0$. However since the sampling of turbulence statistics upon projection is not statistically complete, meaning that the wavevectors with $k>K_0$ but $K<K_0$ is not sampled, it is hard to determine whether we will obtain back the same turbulence spectrum anisotropy just by inspection here since we did have additional knowledge on how the LOS direction is related to the inclination angle. More importantly, if we are considering the case when $M_A$ is not small, the randomness of the magnetic field fluctuation will make the filtering in 2D in Stokes parameters being completely different from that of the 3D. Fig.\ref{fig:HPF} shows an example on how different the Stokes Q look like. On the left of Fig.\ref{fig:HPF} we perform filtering after projection (i.e. 2D), while on the right it is the projection after 3D filtering. We can see that, while the statistical anisotropies for the two maps are roughly the same, the differences of the features are prominent. 

\begin{figure}
\includegraphics[width=0.48\textwidth]{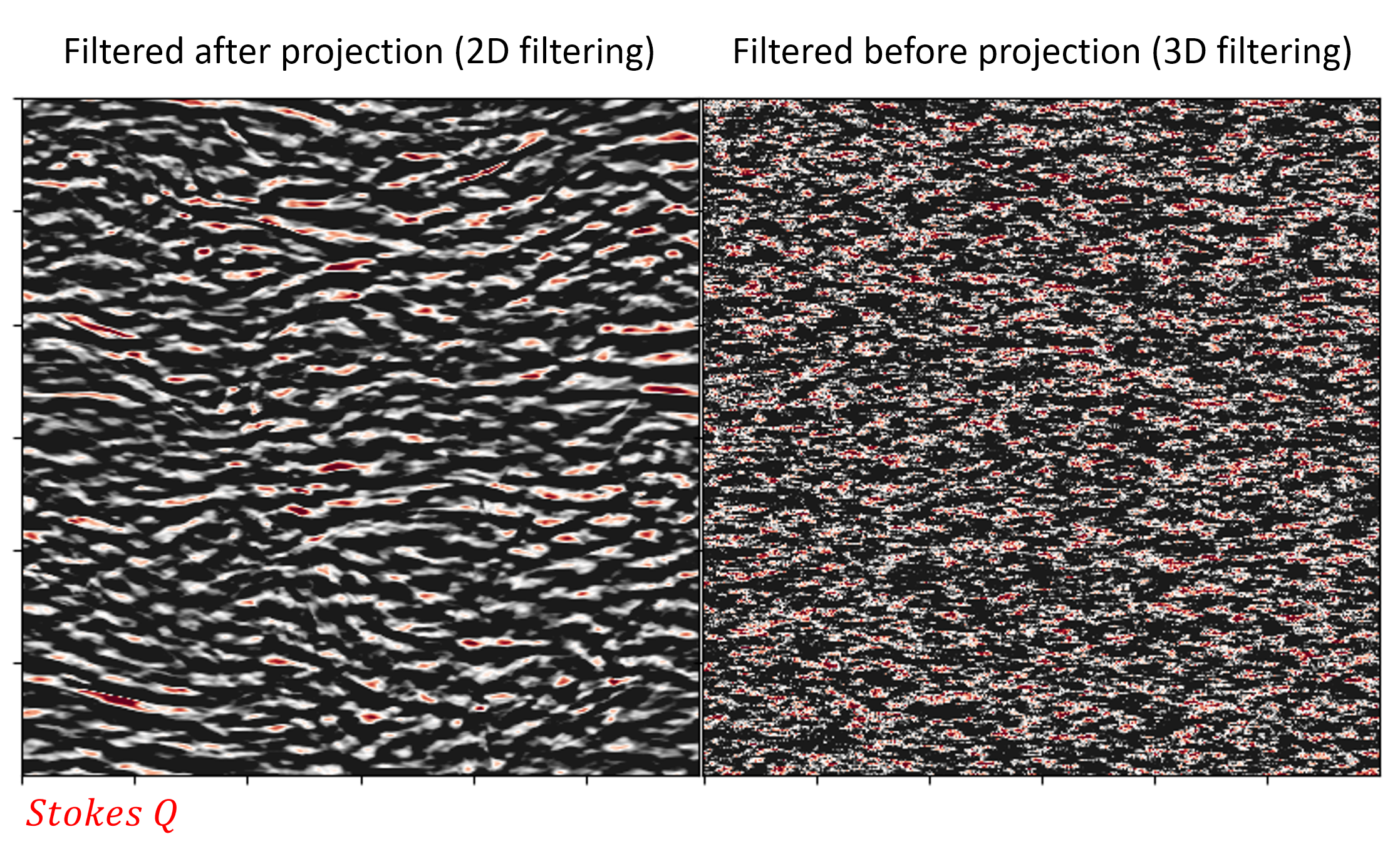}
\caption{An illustration of the features of the Stokes Q after 2D (left) and 3D (right) filtering. One can see that there is a significant difference in terms of the structures of the features. }
\label{fig:HPF}
\end{figure}

}

\section{Conclusion}
\label{sec:conclusion}
In this paper, we introduce a vector-based framework in explaining the strength and the limitation of the recently introduced techniques, namely SPA, CFA and VGT. In particular, due to the use of the vector framework, we recognize that in the presence of curved magnetic field Alfven waves will be seen as the linear combination of Alfven and compressible waves, which is named "Alfven leakage". In short, 
\begin{enumerate}
    \item We recognize a straightforward transformation from the local to global reference frame through the Alfven leakage model. (Fig.\ref{fig:alfleak_incomp}). Moreover, the projection parameters $W_{I,L}$ that are introduced in LP12 are derived in an alternative way in the picture of Alfven leakage. (Eq.\eqref{eq:LP12_WIWL}, Fig.\ref{fig:alfleak_incomp})
    \item The SPA technique, which allows the identifications of the dominance of the Alfven and compressible waves in observed synchrotron emissions, is the result of the one-point statistics. The Alfven wave contribution is frame independent while that for compressible waves are frame dependent. As a result, the quantitative contribution of Alfven and compressible waves can be separated observationally (See Eq.\eqref{eq:dH}).
    \item We suggest that the SPA technique is also applicable to slightly Faraday rotated regime. (\S \ref{sec:SPA_FR}).
    \item Based on the formulation of the Alfven leakage, we discover a new $\gamma$ tracing method that utilize the anisotropy fraction of $I+Q$ and $I-Q$ in observations. We test the method in numerical simulations and see universality of trends across a wide range of turbulence parameters. (\S \ref{sec:gamma}, Fig.\ref{fig:y_model}).
\end{enumerate}
The expression of the vector frame formulation allows us to visually understand and analyze the statistics of MHD turbulence. Together with the theoretical establishment of the \citeauthor{2000ApJ...537..720L} series, how the turbulence statistics are imprinted into observables will be better understood by observers.

\appendix

\section{The mathematical description on vector and tensor formulations in MHD statistical turbulence theory}

For our analysis in this paper, we need to review some of the required mathematical tools for the descriptions of the MHD turbulence. The reason why we need them is because some of the frame representations are advantageous in some situations. Here we will first review the concept of the global and local frame of reference, the leakage of modes due to the \cite{2020ApJ...898...66Y} of local magnetic field, and also the mathematical establishments that are scattered in different literature. The unified approach that we use in this paper will lead to establishment of an analysis framework in understanding how the modes should behave in observations.

\subsection{Global and local frame of reference}

The first important concept is the use of the local frame of reference when computing the structure function of the turbulence variable. The mathematical expression of the 3D structure function  of the turbulence variable $v$ in the local frame of reference is given by:
\begin{equation}
    SF\{{\bf v}\}({\bf r}) = \Big\langle \left(({\bf v}({\bf r'}+{\bf r})-{\bf v}({\bf r'}))\cdot \frac{{\bf B}({\bf r'}+{\bf r})+{\bf B}({\bf r'})}{|{\bf B}({\bf r'}+{\bf r})+{\bf B}({\bf r'})|}\right)^2\Big\rangle
\end{equation}
where in small ${\bf r}$, the separation of the three eigenmodes (Alfven, Fast, Slow) will give the correct spectrum and anisotropy as predicted in GS95 and LV99. In particular, the anisotropy will be scale dependent when observed locally through the 3D structure functions. Table \ref{tab:theory} summarizes the spectral slopes and anisotropies that we expect from the local structure functions.

\begin{table*}
\centering
\begin{tabular}{c c c c}
Mode &  Power spectra $E(k)$, where $E = \int dk E(k)$ & Anisotropy factor & Frame vector \\ \hline \hline
Alfven & $k^{-5/3}$ & $\exp(-M_A^{-4/3} k_\parallel/k_\perp^{2/3}$) & $\zeta_A$\\
Slow (low $\beta$)  & $k^{-5/3}$ & varies (see \citealt{2020PhRvX..10c1021M}) & $\zeta_S$\\
Slow (high $\beta$) &  $k^{-5/3}$  & $\exp(-M_A^{-4/3} k_\parallel/k_\perp^{2/3})$ & $\zeta_S$\\
Fast (low $\beta$)  & $k^{-3/2}$ &  1 & $\zeta_F$\\
Fast (high $\beta$) & $k^{-3/2}$ & $k_\perp^2$ & $\zeta_F$ \\\hline\hline
\end{tabular}
\caption{\label{tab:theory}  A summary of the theoretical expectations of the turbulence scaling laws. Summarized from CL03, Yan \& Lazarian (2008) and Makwana \& Yan (2020).}
\end{table*}

However, we cannot deduce the expressions from Tab.\ref{tab:theory} due to the restriction of the local-to-global reference frame transformation, which is the main topic of the current paper. A more common method in computing the structure function is by simply computing the simplistic structure function below, assuming $V_z({\bf R}) = \int dz \hat{\bf z}\cdot{\bf v}({\bf r}) $:
\begin{equation}
    SF\{V_z\}({\bf R}) = \langle (V_z({\bf R'}+{\bf R})-V_z({\bf R'}))^2\rangle
\end{equation}
which the spectrum and anisotorpy that is observed from this variable could be different from what the local expressions. In particular ,the anisotropy in the global frame of reference becomes scale independent, meaning that there is no particular advantage in probing the anisotropy in smaller scale in actual observations, aside from the standard $LM_A^{-3}$ scale.

\begin{figure*}
\includegraphics[width=0.98\textwidth]{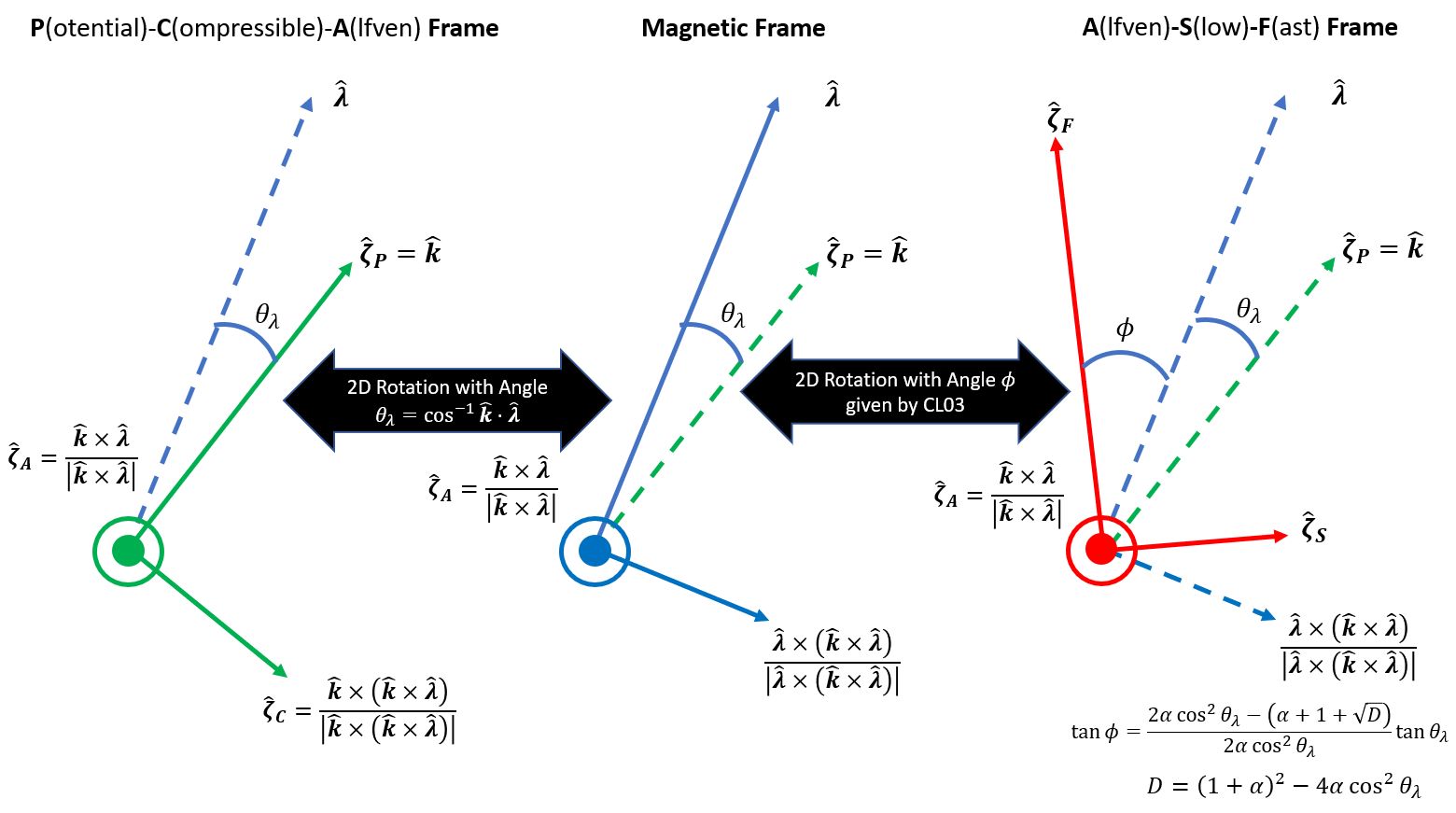}
\caption{The definition of the frames that we use in the current paper. From the left: The Potential-Compressible-Alfven Frame (PCA, left) which is very convenient in analyzing the magnetic field perturbations. The Alfven-Slow-Fast Frame (ASF, right) was the local reference frame for the three fundamental MHD eigenmodes derived in CL03 The magnetic frame (middle) is the frame that shares similarity to the PCA frame defined through the local magnetic field direction $\hat{\lambda}$. To connect them, the PCA frame is a simple rotation of $\theta_{\lambda}=cos^{-1}(\hat{\bf k} \cdot \hat{\bf \lambda})$ from the magnetic frame, and the ASF frame is just a $\phi$ rotation from the magnetic field.  }
\label{fig:illus}
\end{figure*}

.

\subsection{Tensor representation}
\label{subsec:T}
In the global frame of reference, the spectral tensor for different modes can be represented by the sum of the three linearly independent spectral tensors $T_{P,C,A}$, which is given by (\cite{LP12}, cf Yan \& Lazarian 2004):
\begin{equation}
\begin{aligned}
T_{P,ij} &= \hat{k}_i\hat{k}_j\\
T_{C,ij} &= \frac{(\hat{\bf k}\times (\hat{\bf k} \times \hat{\bf  \lambda}))_i(\hat{\bf k}\times (\hat{\bf k} \times {\bf \lambda}))_j}{|\hat{\bf k} \times \hat{\bf  \lambda}|^2}\\
&=\frac{(\lambda_i-(\hat{k}\cdot\hat{\lambda})k_i)(\lambda_j-(\hat{k}\cdot\hat{\lambda})k_j)}{ |\hat{\bf k} \times \hat{\bf  \lambda}|^2}\\
T_{A,ij} &= I_{ij} - T_{P,ij} - T_{C,ij}\\
&= \frac{(\hat{\bf k} \times \hat{\bf  \lambda})_i(\hat{\bf k} \times {\bf \lambda})_j}{|\hat{\bf k} \times \hat{\bf  \lambda}|^2}
\end{aligned}
\end{equation}
Notice that for Alfven mode $v_{A,i} T_{P,ij} = 0$ since $\nabla \cdot v_{A,i}=0$. Notice that $T_{C}+T_{A}$ is isotropic.

\subsection{The ASF (CL03) frame with respect to the PCA frame }

For the actual numerical analysis, the realization of the individual MHD modes in the local frame of reference is not achievable since obtaining the modes requires the perturbation theory to start with. In this case, the expressions of the modes are given in Fourier space by evaluating the perturbation along a locally averaged mean field. In that case, for each ${\bf k} \in \mathcal{R}^3$, we can locally define the eigenvectors for the three modes ${\hat{\zeta}}_{A,S,F}$ given by Eq.\ref{eq:cho}. Notice that the A(lfven)-S(low)-F(ast) frame is a simple rotation of the "magnetic frame" along ${\hat{\zeta}}_A$ given by the three eigenvectors $(\hat{\lambda},\hat{\bf k}\times \hat{\lambda}, \hat{\lambda}\times (\hat{\bf k}\times \hat{\lambda}))$ by an angle $\phi$:
\begin{equation}
    { \tan\phi = \frac{2\alpha\cos^2\theta_{\lambda}-(\alpha+1+\sqrt{D})}{2\alpha \cos^2\theta_{\lambda}}\tan\theta_{\lambda}}
\end{equation}.

The "magnetic field" is simply given by an additional rotation of $\tan\theta_{\lambda}$ from the P(otential)-C(ompressible)-A(lfven) frame $(\hat{\zeta}_P=\hat{\bf k},\hat{\zeta}_A=\hat{\bf k}\times \hat{\lambda}, \hat{\zeta}_C=\hat{\bf k}\times (\hat{\bf k}\times \hat{\lambda}))$. The PCA frame has its special advantage since the sampling of ${\bf k}$ is usually complete in $d\Omega_k$. That means we have the freedom to fix ${\bf k}$ despite other unit vectors are changing.

From the tensor product we can always write the arbitrary vector in the Fourier space as :
\begin{equation}
    \zeta_i({\bf k}) = C_P \hat{k}_i + C_C \frac{(\hat{\bf k}\times (\hat{\bf k} \times \hat{\bf  \lambda}))_i}{ |\hat{\bf k} \times \hat{\bf  \lambda}|} + C_A \frac{(\hat{\bf k} \times \hat{\bf  \lambda})_i}{|\hat{\bf k} \times \hat{\bf  \lambda}|}
\end{equation}
which we will name the unit vector $\zeta_{P,C,A}$ now

From \cite{2003MNRAS.345..325C}, in the global frame of reference the Alfven, slow and fast mode eigenvectors are:
\begin{equation}
    \begin{aligned}
    \zeta_A &\propto \hat{\bf k} \times \hat{\bf  \lambda}\\
    \zeta_S &\propto (-1 +\alpha-\sqrt{D}) ({\bf k}\cdot \hat{\bf \lambda}) \hat{\bf \lambda}  + (1+\alpha - \sqrt{D}) ( \hat{\bf \lambda} \times ({\bf k}\times \hat{\bf \lambda})) \\
    \zeta_F &\propto (-1 +\alpha+\sqrt{D}) ({\bf k}\cdot \hat{\bf \lambda}) \hat{\bf \lambda}  + (1+\alpha + \sqrt{D}) ( \hat{\bf \lambda} \times ({\bf k}\times \hat{\bf \lambda})) \\
    \end{aligned}
    \label{eq:cho2}
\end{equation}
where $\alpha = \beta\gamma/2$, $D=(1+\alpha)^2- 4\alpha\cos ^2\theta$, $\cos\theta = \hat{\bf k}\cdot \hat{\bf \lambda}$. We recognize that there is a frame rotation between the vector $\zeta_{P,C}$ and $\zeta_{S,F}$:
\begin{equation}
    \begin{bmatrix}
    \zeta_S\\ \zeta_F
    \end{bmatrix} = \frac{-1}{2\cos2\theta \sqrt{D}}{\bf L}(\alpha,\theta) {\bf R_0}(\theta)
     \begin{bmatrix}
    \zeta_P\\ \zeta_C
    \end{bmatrix}
    \label{eq:Cho_to_chepurnov}
\end{equation}
where ${\bf R_0}(\theta)$ is the standard two-dimensional rotation matrix, the factor beforehand is just for normalization and: 
\begin{equation}
\begin{aligned}
    {\bf L} (\alpha,\theta) &= 
    \begin{bmatrix}
    (-1 +\alpha-\sqrt{D})\cos\theta  & (1+\alpha - \sqrt{D})\sin\theta \\
    (-1 +\alpha+\sqrt{D})\cos\theta & (1+\alpha + \sqrt{D})\sin\theta 
    \end{bmatrix}
\end{aligned}
\end{equation}

Then we can rewrite the tensors by 
\begin{equation}
    \begin{aligned}
    T_{S/F} = \zeta_{S/F}\otimes\zeta_{S/F}
    \end{aligned}
\end{equation}
Notice that $T_{ij}\zeta_j = \zeta_i$ if $T_{ij} = \zeta_i\otimes\zeta_j$.

\subsection{Frenet-Serret frame}

From \cite{2020ApJ...898...66Y}, the Frenet-Serret frame of the the magnetic fields lines would be:
\begin{equation}
    \begin{aligned}
    \frac{d\hat{t}}{ds} &= &+\kappa \hat{n}&\\
    \frac{d\hat{n}}{ds} &= -\kappa \hat{t}&&+\tau \hat{b}\\
    \frac{d\hat{b}}{ds} &= &-\tau \hat{n}&\\
    \end{aligned}
    \label{eq:FSF}
\end{equation}
Here $\hat{t} = \hat{\lambda}$, representing the tangent vector of the magnetic field line. $(\hat{t},\hat{n},\hat{b})$ forms a complete orthogonal set independent of the choice of $k$. Notice that for mode decomposition, the "mean" field is selected before selecting (Fourier transforming into) ${\bf k}$, thus we can treat ${\bf\lambda}$ as k-independent and uses its own position vector ${\bf r_\lambda})$. Notice that the unit vector $\hat{n}$ can be expressed as the linear combination of ${\hat{\zeta}}_A$ and $\hat{\lambda}\times {\hat{\zeta}}_A$ in the magnetic frame

\subsection{The relation between the tensor representation (Lazarian \& Pogosyan 2012) and vector representation (this work)}

In the local frame of reference, the Alfven mode magnetic field is given by simply:
\begin{equation}
    {\bf H}_A({\bf r}) = \int d^3 k C(k) \hat{\zeta}_A(k)
\end{equation}
where $C$ contains the isotropic and anisotropic factors from its spectrum. However as we move from the local frame to the global frame, the actual Alfven wave magnetic field realization will contain both compressible and Alfven wave contribution (here we simply pick an arbitrary ${\bf k}$):
\begin{equation}
    \tilde{H}_A({\bf k}) = C W_A \hat{\zeta}_A + C W_C \hat{\zeta}_C
\end{equation}
where $W_{A,C}$ are two factors yet to be found. LP12 branded these two factors in the form of the direct tensor product $\zeta_{E} = \zeta_{C}+\zeta_{A}$ and $\zeta_{F} = \zeta_{C}$, and $T_{E,F} = \hat{\zeta}_{E,F} \otimes \hat{\zeta}_{E,F}$. In their case when Alfven mode is observed in the local frame of reference, the Alfven mode correlation function in k-space is given by:
\begin{equation}
    \tilde{H}_i \tilde{H}_j = C^2 (T_{E,ij}-T_{F,ij})
\end{equation}
while in the global frame of reference
\begin{equation}
    \tilde{H}_i \tilde{H}_j = C^2 T_{E,ij}-C^2(W_I T_{E,ij} + W_LT_{F,ij})
\end{equation}
Some algebra will give
\begin{equation}
    \tilde{H}_i \tilde{H}_j = C^2 (1-W_I-W_L) T_{C,ij}+ C^2(1-W_I)T_{A,ij}
\end{equation}

\subsection{Conversion between the frame of references of velocity field and magnetic field}

{ As derived by \cite{2002PhRvL..88x5001C,2003MNRAS.345..325C} the decomposed Alfven-Slow-Fast frame was the frame for the displacement vector $\zeta$, which also applies to the velocity fluctuations. However the magnetic field fluctuations do not necessary follow the ASF frame as defined in CL03. For Alfven wave, the fluctuations of the magnetic field is in the same direction as that of velocities, i.e. $\hat{k}\times\hat{\lambda}$. For compressible modes, the propagation of the magnetic field fluctuations $\tilde{b}({\bf k})$ at a specific wavevector $\bf k$ is given by the following relation:
\begin{equation}
    \tilde{b} = \hat{k} \times (\tilde{v} \times \hat{\lambda})
\end{equation}
where $\tilde{v}$ is the velocity fluctuation at $\bf k$. Notice that the above vector is parallel to the compressible vector $\hat{\zeta}_C = \hat{\bf k}\times (\hat{\bf k} \times \hat{\bf  \lambda})$.
}

\section{The fundamentals of describing the anisotropy in structure functions}
\label{ap:anisotropy}

In this section we will discuss the essence of multipole expansions in analysing the statistics of turbulence under the assumption of two-point closure\footnote{The concept of two-point closure is simply to say that turbulence variables can be "adequately" described by the two-point structure functions. This approximation is evidently incorrect in general turbulence case as intermittency is a well-studied topic in the field. However for equilibrium MHD turbulence that we are considering here, the two-point description contains $\sim 95\%$ of the spectral power. The prominent features that we are measuring (e.g. mode fraction, $\gamma$ etc) are therefore dominated by the two-point statistics. See  Yuen (thesis, 2022) } based on the formalism of \cite{KLP16}.

It is visually compelling that the two-point structure functions are concentric ellipses. Mathematically the structure functions of anisotropic fundamental modes (e.g. Alfven, slow modes) contains a dependence in the form of $\exp(-C |\cos\phi|)$ for some constant $C$ that carries a weak dependence on $\phi$ (See, e.g. \citealt{LP12,KLP16}). This exponent term is naturally elliptical like. The expression of this term in the two-point statistics of any observables is the main direction of theoretical study recently in literature \citep{LP12,LP16,KLP16,KLP17,ch5,ch9}.

\begin{figure*}
\includegraphics[width=0.98\textwidth]{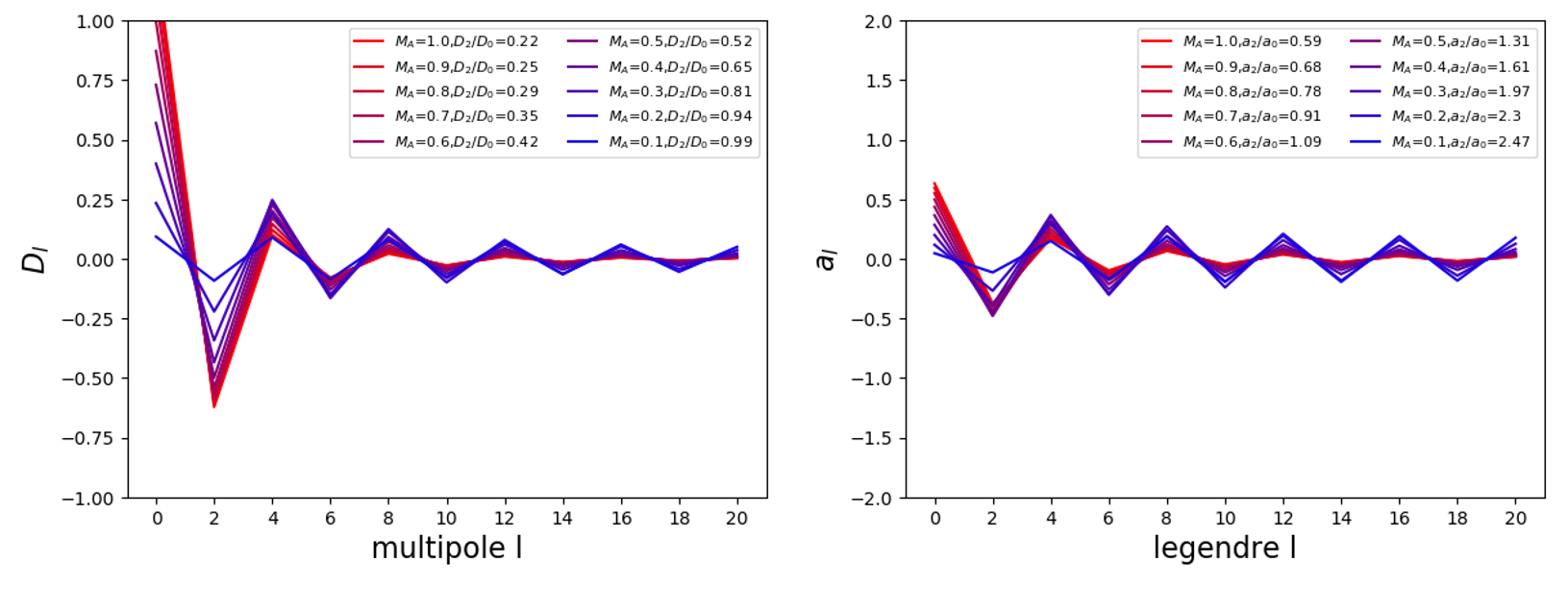}
\caption{Two figures showing how the change of $M_A$ affects the relative amplitude of $D_l$ and $a_l$. We can see from the L.H.S. figure that, when $M_A$ is large, the approximation til quadruple is very good since $|D_4|<min(D_2,D_0)$. However as $M_A$ becomes smaller, $|D_4|$ is actually comparable to that of $|D_{0,2}|$, and therefore the multipole approximation breaks down . Very similar result also happens for Legendre expansion (right).}.
\label{fig:multipole2}
\end{figure*}

There are a few choices in describing elliptical features on the sky via complete basis:\linebreak\linebreak
\noindent {\bf Multipole expansions of even order}: The spatial symmetry of the function $\exp(-C |\cos\phi|)$  allows one to express the structure function of any observables $X$ into the summation the cosines with even orders: $D_{X}(\phi) =  \sum_{m\in 	2\mathbb{Z}^+} D_{m}\cos(m\phi)$ \footnote{In previous literature (e.g. \citealt{KLP16}) they express $D_{X}\sim \sum_{m\in 	2\mathbb{Z}} \bar{D}_m e^{im\phi}$, where $m$ can be both positive and negative. Typically structure functions are always real-valued. Therefore for the sake of simplicity we adopt the cosine formalism.}. Visually we are expressing the structure function into linear combination of cosines in polar coordinate. Notice that for all $m|4=2$ the $D_m$ term carries some anisotropy, however for $m\ge 6$ the multipole anisotropy has a upper limit. For instance, the $\cos 6\phi$  term has a maximum anisotropy of 1.15. Notice that the non-vanishing $D_{m\ge6}$ will {\it decrease} the anisotropy of the structure function. A typical treatment of the multipole expansion is to truncate the series into $m=0,2$, where the visual minor-to-major axis ratio for the elliptical feature appeared in the structure function $\chi = \sqrt{1-\epsilon_{ell}^2}$ ($\epsilon_{ell}$ is the eccentricity of ellipse) is given by:
\begin{equation}
\chi = \frac{D_0-D_2}{D_0+D_2}
\end{equation}
Notice that the multipole expansion fails when $M_A\ll 1$ or $M_A>1$, as the $D_m$ term is comparable $\forall m$. The empirical limit where $D_4/D_2$ is comparable ($\sim 0.5$,\citealt{ch9}) is roughly at $M_A\approx 0.5$. Therefore the multipole expansion is suitable only for $M_A \sim 0.5-1$ (See Fig.\ref{fig:multipole2})
\linebreak\linebreak
\noindent {\bf Legendre Polynomial}: The Legendre polynomial $P_l(\cos\phi$ is another popular choice in describing the statistics in 2D. Similar to the multipole expansion, we express the structure function $D_X(\phi)=  \sum_{l\in 	2\mathbb{Z}^+} a_{l}P_l$. $a_{l}$ carries very similar mathematical properties as $D_m$ in multipole expansions and therefore we would not discuss further. (See Fig.\ref{fig:multipole2})
\linebreak\linebreak

\noindent{\bf Elliptical basis}: As the structure function look like ellipses, it is natural to consider the function below to capture the anisotropy of the structure function:
\begin{equation}
f(\phi,\epsilon_{ell}) = \frac{\sqrt{1-\epsilon_{ell}^2}}{(1-\frac{\epsilon_{ell}^2}{2} + \frac{\epsilon_{ell}^2}{2} \cos 2\phi)}
\label{eq:ellipse}
\end{equation}
The advantage of this basis is that (1) the eccentricity $\epsilon_{ell}$ is a direct measure of the minor-to-major axis ratio, which allows one to quickly construct this function by simply measuring the minor and major axis (2) due to the non-vanishing higher-order multipole of Eq.\ref{eq:ellipse}, this functional form is still applicable when $M_A\ll1$. Notice that one can convert the eccentricity $\epsilon_{ell}$ to the $D_2/D_0$ via the formula:
\begin{equation}
\Big|\frac{D_2}{D_0}\Big|\approx \frac{1}{2} \frac{2\epsilon_{ell}^2}{2-\epsilon_{ell}^2}
\label{eq:d2d0_to_e}
\end{equation}
in which the approximation is valid when $M_A\in [0.5,1]$ {\it for the case of linear} (i.e. centroid, $C \propto \int dz v_z$) or {\it quadratically projected observables (i.e. Stokes parameters). }. The approximation is valid for caustics (c.f. \citealt{VDA}) for even smaller values of $M_A$.

{\noindent\bf Acknowledgments.}    K.H.Y. \& A.L. acknowledge the support the NSF AST 1816234, NASA TCAN 144AAG1967 and NASA ATP AAH7546. KHY thanks Dmitri Pogosyan (U.Alberta) and Ka Wai Ho (UW-Madison) for their inspirational comments. We thank Sunil Malik (DESY) and Parth Pavaskar (DESY) for extensive discussions and cross-checks on the validity of the y-parameter analysis. The main simulations and the first version of the work is done during KHY's tenure in UW Madison. Research presented in this article was supported by the Laboratory Directed Research and Development program of Los Alamos National Laboratory under project number(s) 20220700PRD1.

{\noindent \bf Code Availability} The code can be found in \url{https://github.com/kyuen2/MHD_modes}

{\noindent \bf Data Availability} The data underlying this article will be shared on reasonable request to the corresponding author. 

\bibliographystyle{mnras}
\bibliography{refs} 

\bsp	
\label{lastpage}

\end{document}